\documentclass{article}
\usepackage[utf8]{inputenc}
\usepackage[style=nature,sorting=none]{biblatex}
\bibliography{ref}
\renewbibmacro{in:}{%
  \ifentrytype{article}
    {}
    {\bibstring{in}%
     \printunit{\intitlepunct}}}
\DeclareFieldFormat[article]{pages}{#1}
\usepackage{amsmath,amsthm,amssymb,multicol,xcolor}
\usepackage{bbm}
\usepackage{graphicx}
\usepackage[margin=20truemm]{geometry}
\usepackage{istgame}
\usepackage{subcaption}
\usepackage{caption}
\usepackage{dsfont}
\usepackage{threeparttable}
\usepackage{cancel}
\usepackage{float}
\usepackage{booktabs}
\usepackage{algorithm}
\usepackage{algpseudocode}
\usepackage{authblk}
\usepackage{tikz}
\usepackage{tikz-qtree}
\usetikzlibrary{trees, arrows.meta, positioning, shapes, calc,backgrounds}
\usepackage[edges]{forest} 

\theoremstyle{plain}
\newtheorem*{theorem*}{Theorem}

\renewcommand{\thesubsection}{\arabic{subsection}}

\title{The Evolution of Trust under Institutional Moral Hazard}
\date{}
\author[*]{Hiroaki Chiba-Okabe}
\author[*]{Joshua B.~Plotkin}
\affil[*]{University of Pennsylvania}

\begin{document}

\maketitle

\begin{abstract}
\noindent We study the behavior of for-profit institutions that broadcast reputations to foster trust among market participants. We develop a theoretical model in which buyers and sellers are matched on a platform to engage in transactions involving a moral hazard: sellers can either faithfully deliver goods after receiving payment, or not.  Although the buyer does not know a seller's true type, the platform maintains a reputation system that probabilistically assigns binary reputation signals. Buyers make 
purchase decisions based on reputation signals, which influence the payoffs to sellers who then adapt their type over time. These market dynamics ultimately shape the platform’s profit from commissions on sales. Our analysis reveals that platforms inherently have an incentive for rating inflation, driven by the desire to increase commission. This introduces a second layer of moral hazard: the platform’s incentive to distort reputations for its own profit. Such distortion is self-limited by the platform's need to maintain enough accuracy that trustworthy sellers remain in the market, without which rational buyers would refrain from purchases altogether. Nonetheless, the optimal strategy for the platform can be to invest in order to reduce signal accuracy. When the platform can freely set commission fees, however, maximum profit may be achieved by costly investment in an accurate reputation system. These findings highlight the intricate tensions between platform incentives and resulting social utility for marketplace participants.
\end{abstract}

\section{Introduction}

Institutions play a significant role in promoting moral and cooperative behavior in human societies \cite{zucker1986,cook2005,Greif_2006,ostrom2010}. This can be done not only by directly modifying payoffs to individuals \cite{gurek2006,kosfeld2009,schoenmakers2014,sigmund2010,sigmund2011,dong2019,aagren2019enforcement,duong2021,Lim2024Trust,chibaokabe2024,liu2025}, but also by transmitting information about social reputations \cite{radzvilavicius2021,kessinger2024institutionspublicjudgmentestablished,liepanis2024}, which is recognized as a key driver of cooperation \cite{nowak1998,leimar2001,milinski2002,fu2008}. Historically,  institutions that monitor and broadcast reputations were instrumental in facilitating trade by generating trust among strangers \cite{milgrom1990,greif2004impersonal}. In the 19th-century development of America, for example, credit bureaus were a prerequisite for the expansion of industry across wide geographic expanses \cite{carruthers2022economy}. 

Institutions that monitor and broadcast reputations continue to play an important role in today's economy, especially online. Prominent examples are online transaction platforms, which often display ratings about the quality and trustworthiness of market participants \cite{resnick2002trust}. Such a reputation system can significantly impact the dynamics of interactions on the platform by building trust among market participants \cite{resnick2000reputation,tadelis2016}. While a public rating system may tend to promote good-faith behavior among its participants \cite{dellarocas2005,benson2020,liu2021}, it is unclear whether the platform itself is incentivized to report reputations honestly or not. Online marketplace platforms are managed by for-profit private companies who typically receive commissions from sales made on their platforms \cite{hasiloglu2021}, and so their interests may not align with those of all users. Platforms have the ability to set policies that shape the informational structure and content within these marketplaces, thereby influencing participant behavior and market dynamics. Although the rating systems often rely on post-transaction feedback generated by the users, the platform can choose whether to disclose the evaluations faithfully; and they can choose whether or not to police and remove fake user reviews \cite{wu2020}. This substantial level of control gives a platform the ability to manipulate or distort reputational information to prioritize its own interests, potentially undermining the integrity and efficiency of the marketplace.

In this study, we develop theoretical models to examine the role of a for-profit reputation institution, such as an online transaction platform, in fostering cooperative behavior among players of a trust game. The trust game captures the essential structure of market transactions with the possibility of moral hazard \cite{dellarocas2005,tadelis2016}. Our analysis will account for both the incentives of the institution itself as well as the incentives of the buyers and sellers engaged in the trust game on the platform. In the literature on social evolution, theoretical models have been used to study how institutions can incentivize cooperation by assigning rewards \cite{Lim2024Trust,liu2025} and reputations \cite{bravo2008,hu2021,liu2023}; but the incentives of the institutions themselves have received less attention. Although the theory for the evolution of cooperation has begun to explore the incentives of institutions, most studies have focused on the Prisoner's Dilemma, which has a symmetric structure in which players act simultaneously \cite{kessinger2024institutionspublicjudgmentestablished,chibaokabe2024}; whereas the trust game has an asymmetric, sequential structure in which the first actor must choose whether to trust the second actor, who may or may not reciprocate. 

This study is also related to the extensive literature on online transaction platforms.  Theoretical models have examined the effects of product reviews on seller strategies \cite{kwark2014,kwark2017}, the impact of review manipulation \cite{dellarocas2006,mayzlin2006}, and platform incentives related to search costs \cite{duke2016,jiang2020} and the extent of information disclosure \cite{li2017,shi2023}. Our study departs from this literature by incorporating a social evolution perspective, whereby sellers adaptively switch trustworthy and untrustworthy types in response to the platform's policies, as well as by allowing the platform to actively distort reputation information. Although an evolutionary model has been applied to study the coevolution of sellers' and buyers' behavior \cite{masuda2012}, the possibility for the platform to manipulate the reputations to its advantage has not been studied. Modeling both the behavioral evolution of market participants and the institution's policies allows us to capture a realistic dynamic where market participants adapt to the environment created by the platform, which in turn affects the incentives of the platform itself. 

We develop a model for simple setting in which the institution, or platform, establishes a binary reputation system that displays the reputations of sellers, who may be good or bad. Good sellers are trustworthy and deliver the purchased good or service, whereas bad sellers are untrustworthy---they fail to deliver. A buyer chooses whether or not to make a purchase from a seller after seeing the reputation assigned by the platform, aiming to maximize its expected payoff. The platform receives commissions from sellers for each sale they make, which corresponds to standard practice in online transaction platforms \cite{hasiloglu2021}. To investigate the interplay between the platform incentives and market dynamics, we allow the platform to control the accuracy of the reputation system, possibly incurring higher cost to achieve higher accuracy; and we study the behavioral dynamics of sellers who switch their types by payoff-biased imitation.

As we will demonstrate, our model produces a unique equilibrium in which the platform earns a strictly positive payoff. In this equilibrium, good and bad sellers coexist, and their frequencies depend on the accuracy of the reputation system provided by the platform. When it is cost-free for the platform to set any accuracy, we show that the platform maximizes its revenue by being perfectly accurate about the reputations of good sellers, meaning that good sellers are assigned a ``good'' signal with probability $1$. 
The platform can still maximize profit while allowing bad sellers to be assigned the good label sometimes; but there is an upper bound on the probability of assigning the ``good'' signal to sellers that, in truth, are bad. The problem of moral hazard by the platform surfaces when maintaining an accurate reputation system becomes costly. The platform has little incentive to improve the accuracy of the reputation system, and it may even be optimal for the platform to artificially reduce the accuracy of signals so that it becomes harder to distinguish between good and bad sellers. This indicates systematic incentives for the platform to be lenient toward, or to cause, rating inflation---a phenomenon where ratings are artificially elevated, which is observed in real-world online transaction platforms \cite{nosko2015,zervas2020airbnb,filippas2022}. These results arise from the platform's mixed incentives: to keep the equilibrium share of good sellers high while facilitating transactions with bad sellers that necessarily exist in the equilibrium. Our analysis highlights the potential conflict of interest between a platform and its users, since buyers and good sellers enjoy higher utility when the reputation system is more accurate. When the platform can freely choose the commission fee---such as when it holds a dominant position that leaves sellers with few viable alternatives---commission rates introduce another layer of strategic complexity: higher fees increase per-transaction revenue but necessitate a more accurate reputation system to retain good sellers with thinner profit margins. This creates a trade-off, as the platform must balance its revenue objectives against the risk of driving away high-quality sellers.

This study contributes to the literature of the evolution of trust under institutions \cite{hu2021,liu2023,Lim2024Trust} by clarifying the conditions under which reputational institutions with their own incentives can benefit the society. By doing so, we also extend studies on the interplay between the evolution of cooperation and institutional incentives \cite{chibaokabe2024,kessinger2024institutionspublicjudgmentestablished}. Our findings are also relevant to the literature on e-commerce platforms, especially concerning rating inflation. For example, it has been suggested that rating inflation might be the result of incentives of the market participants, rather than the platform, such as self-selection \cite{li2008}, reciprocity \cite{dellarocas2008,bolton2013,nosko2015,fradkin2021}, altruism \cite{filippas2022}, social influence \cite{muchnik2013,wang2018} or through creation of fake reviews \cite{brown2006,mayzlin2014,xu2015}. Whereas our analysis shows that rating inflation may be caused by platform incentives alone, which largely agrees with empirical evidence that inflation  boosts sales \cite{aziz2023}. Our account therefore provides a framework for studying what incentives will cause platforms to compromise their rating accuracy, to the detriment of good-faith users.

\section{Model}

We assume a single marketplace platform and an infinite population of sellers. Each seller belongs to one of three types: good ($G$), bad ($B$), or inactive ($I$). We denote the share of each type of sellers within the population by $x_{i}, i\in\{G,B,I\}$. Good sellers faithfully deliver goods or services, whereas bad sellers fail to do so. Inactive sellers opt out ex ante from the platform and do not engage in any transaction. The platform assigns active sellers a signal $\hat{G}$ with probability $\alpha$, if the true type of the seller is $G$, and with probability $\beta$ if the true type of the seller is $B$. Otherwise the signal $\hat{B}$ is assigned. This probabilistic formulation reflects the noisy characteristic of reputation systems \cite{ranchordas2018online}. We refer to $\alpha$ as the true positive rate, and $\beta$ the false positive rate, of the platform's reputation system. The reputation system is perfectly accurate when $\alpha=1$, and $\beta=0$.

A buyer visits the platform at some rate and is matched randomly with one of the active sellers. Seeing the signal assigned to the seller by the platform, the buyer chooses its action (purchase or not) to maximize its expected payoff. If the buyer decides to purchase and the true type of the seller is $G$, both the seller and the buyer gain payoff $r>0$, which we call the benefit of good-faith transactions. On the other hand, if the seller was bad, the seller gains $1$ and the buyer experiences payoff $-1$. If no purchase is made, payoffs are $0$ for both buyer and seller. We restrict our attention to the case where $r<1$, in line with the typical formulation of the trust game. This payoff structure creates a social dilemma where it is individually rational for the sellers to not reciprocate, which in turn, incentivizes buyers not to trust sellers. Without any reputation system or incentive schemes, this results in a trivial outcome where no transactions occur \cite{Lim2024Trust}. Aside from the payoff they gain from the transaction itself, sellers also pay commission $c>0$ to the platform for each sale they make.

\begin{figure}
\centering
\begin{tikzpicture}[
  every path/.style={thick},
  round node/.style={fill=black,circle,draw,inner sep=1},
  level distance=10mm,
  level 1/.style={sibling distance=57mm},  
  level 2/.style={sibling distance=28mm},  
  level 3/.style={sibling distance=15mm},  
  font=\footnotesize
]

\node[round node] (root) {}
  child {node[round node] (G) {} 
    child {node[round node] (Gg) {} 
      child {node[round node,label=below:{$(r,r)$}] {}
        edge from parent node[left] {Buy}}
      child {node[round node,label=below:{$(0,0)$}] {}
        edge from parent node[right] {Not buy}}
      edge from parent node[above left, xshift=-1mm, yshift=-2mm] {$\hat{G}$}
    }
    child {node[round node] (Gb) {} 
      child {node[round node,label=below:{$(r,r)$}] {}
        edge from parent node[left] {Buy}}
      child {node[round node,label=below:{$(0,0)$}] {}
        edge from parent node[right] {Not buy}}
      edge from parent node[above right, xshift=1mm, yshift=-2mm] {$\hat{B}$}
    }
    edge from parent node[left, xshift=1mm,yshift=2mm] {$G$}
  }
  child {node[round node] (B) {} 
    child {node[round node] (Bg) {} 
      child {node[round node,label=below:{$(-1,1)$}] {}
        edge from parent node[left] {Buy}}
      child {node[round node,label=below:{$(0,0)$}] {}
        edge from parent node[right] {Not buy}}
      edge from parent node[above left, xshift=-1mm, yshift=-2mm] {$\hat{G}$}
    }
    child {node[round node] (Bb) {} 
      child {node[round node,label=below:{$(-1,1)$}] {}
        edge from parent node[left] {Buy}}
      child {node[round node,label=below:{$(0,0)$}] {}
        edge from parent node[right] {Not buy}}
      edge from parent node[above right, xshift=1mm, yshift=-2mm] {$\hat{B}$}
    }
    edge from parent node[right, xshift=-1mm,yshift=2mm] {$B$}
  };

\draw[dashed, bend left, gray] (Gg) to (Bg);  
\draw[dashed, bend right, gray] (Gb) to (Bb); 

\coordinate (legendX) at ($(Bb)!0.5!(root) + (-8.5,0)$);
\node at (legendX|-root)   {Seller type};
\node at (legendX|-Gb)     {Purchase decision};
\node at (legendX)         {Reputation signal};

\end{tikzpicture}
\caption{\small An extensive-form trust game between buyer and seller, with incomplete information. Nature assigns the seller type and the reputation signal, and the buyer makes a purchase decision without observing the seller's true type. The dashed lines represent the buyer’s information sets, indicating that the buyer cannot distinguish the true type given a reputation signal. Terminal nodes show payoffs in the order (buyer, seller).}
\label{tree}
\end{figure}
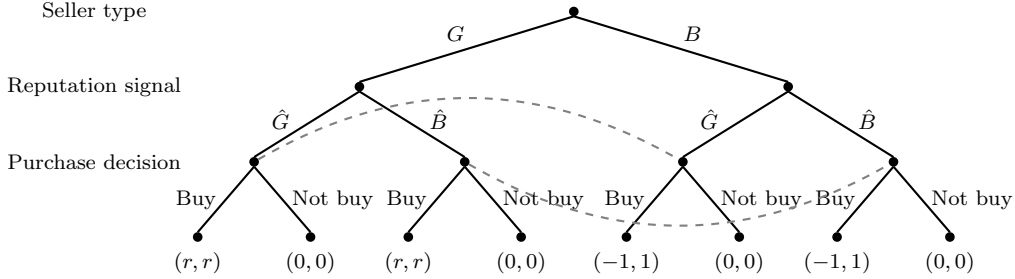

According to this model formulation, each buyer 
engages in a form of trust game with incomplete information, similar to previous models of online transaction platforms \cite{dellarocas2005,masuda2012,tadelis2016}. Specifically, it is an extensive form game consisting of three stages: a seller type, either $G$ or $B$, is assigned by nature; a signal, either $\hat{G}$ or $\hat{B}$, is assigned to the matched seller by  nature; having observed the signal (but not the true type of the seller), the buyer decides to purchase or not (Figure~\ref{tree}). In this final stage, the buyer chooses the action that maximizes its expected payoff. Specifically, assuming that $\alpha$, $\beta$ and $\xi:=x_G/(x_G+x_B)$, the frequency of good among active sellers are known, the buyer computes the expected payoff in a Bayesian manner and makes a purchase provided the expected utility from purchasing is greater than $0$, which is the payoff of not purchasing. The buyer will not purchase if the expected payoff from purchasing is negative; and the buyer flips a fair coin when the expected utility is exactly $0$.

We suppose sellers update their types (good or bad) according to a classic model of behavioral evolution by payoff-biased imitation,  so that the type frequencies are described by the replicator equation \cite{taylor1978,schuster1983replicator,cressman2014}. This presupposes a gradual process in which sellers change their general attitude rather than directly computing the payoff-maximizing action for each transaction; and is suitable for describing behaviors with moral and normative consequences \cite{McKenzie2007,harms2008,Young2015}. Specifically, the frequencies of different types of sellers evolve over time according to the ordinary differential equation:
\begin{equation}
\dot{x}_{i}=x_{i}\left(\pi_{i}-\sum_{j}x_{j}\pi_{j}\right)
\label{eq:replicator}
\end{equation}
where $x_{i}$ and $\pi_{i}$ are the share and payoff of type $i$, respectively. In our model, the payoffs of sellers with type $G$ and $B$ are their expected per-transaction payoffs:
\begin{equation*}
\begin{split}
\pi_{G}&\triangleq\alpha \pi_{G}^{\hat{G}}+(1-\alpha)\pi_{G}^{\hat{B}}\\
\pi_{B}&\triangleq\beta \pi_{B}^{\hat{G}}+(1-\beta)\pi_{B}^{\hat{B}}\\
\pi_{I}&\triangleq 0
\end{split}
\end{equation*}
where $\pi^{\hat{G}}_{G}$ is the payoff of a good seller who is matched and has been assigned signal $\hat{G}$, and so on. Since inactive sellers never participate in the game, their payoffs are assumed to be zero. This system of differential equations can be seen as an approximation of a stochastic system of a finite number of sellers, where the sellers imitate more successful peers as observed in human decision-making \cite{APESTEGUIA2007217,grujic2020}, in the appropriate limit (see Supplementary Information).

We define the platform revenue to be the product of the per-transaction commission $c$ and total proportion of successful sales, at the equilibrium of the dynamics. There is also cost $C$ associated with the values of $\alpha$, $\beta$, which describes the cost of learning and reporting seller types accurately. The platform's profit, $\Pi$, is then:
\begin{equation*}
\begin{split}
\Pi &\triangleq c\left(\left[\alpha p^{\hat{G}}_{G}+(1-\alpha)p^{\hat{B}}_{G}\right] x^{\ast}_{G}+\left[\beta p^{\hat{G}}_{B}+(1-\beta)p^{\hat{B}}_{B}\right]x^{\ast}_{B}\right)-C(\alpha,\beta)
\end{split}
\end{equation*}
where $x^{\ast}_{G},x^{\ast}_{B}$ are the equilibrium share of good and bad sellers, and $p^{\hat{G}}_{G}\in\{0,1/2,1\}$ is the probability of good sellers assigned with signal $\hat{G}$ successfully making a sale, and so forth. 

Although it is not a priori clear what exact form for the cost function $C(\alpha,\beta)$ is realistic. But it is reasonable to assume that larger costs are incurred as the platform tries to move away from some baseline true- and false-positive rates $(\alpha,\beta)=(\alpha_{0},\beta_{0})$, which can be thought of as the natural accuracy of the signal. For example, signals could be assigned based on a simple coin flip irrespective of the true type (i.e. $\alpha_{0}=\beta_{0}$=1/2), which is not particularly informative. Alternatively, the natural accuracy of the rating system might be achieved by setting up a simple rating system that honestly reports aggregate user feedback, without making any additional efforts to improve accuracy. In this situation,  it is quite possible that the baseline signaling rates will satisfy  $\alpha_{0}>\beta_{0}$: good sellers will more likely be assigned the $\hat{G}$ signal more often than bad sellers. But it is unlikely in practice that $\alpha_{0}=1,\beta_{0}=0$, corresponding to perfect natural accuracy, due to the possibility of sellers creating fake reviews to boost their own rating or harm rivals' reputations \cite{brown2006,mayzlin2014,xu2015}.

\section*{Results}
\setcounter{subsection}{0}

\subsection*{Long-Run Equilibrium of the Market}

We focus on results when the true-positive rate exceeds the false-positive rate, $\alpha>\beta$. Although signals are labeled $\hat{G}$ and $\hat{B}$ for convenience, they are arbitrary binary signals, and the results for cases when $\alpha<\beta$ directly follow by flipping the roles of the signals (or redefining the probabilities as $\alpha'\triangleq 1-\alpha$ and $\beta'\triangleq 1-\beta$). Importantly, if $\alpha(r-c)>\beta(1-c)$ is satisfied, then there is a unique asymptotically stable equilibrium for the evolution of seller types; and good and bad sellers coexist at the equilibrium. The frequency of good sellers in this equilibrium is given by
\begin{equation*}
\xi^{\ast} = \frac{1-\beta}{r(1-\alpha)+(1-\beta)},
\end{equation*}
which accords with the intuition that the equilibrium share of good sellers increases when the true-positive rate $\alpha$ is higher and false-positive rate $\beta$ lower, corresponding to more accurate signals (see also Methods).

The unique stable equilibrium arises from the interplay between buyer decision-making and seller behavioral adaptation. Consider first a situation where most active sellers in the population are good. Such a population cannot be stable because, if only a minute portion of sellers in the market are bad, buyers would purchase from any seller, irrespective of the signal, as the expected payoff is high in such environment. Bad sellers would then enjoy an advantage, earning greater profit than good sellers. This allows invasion and growth of bad sellers until the share of good sellers is pushed down to the equilibrium value $\xi^{\ast}$. On the other hand, when the share of good sellers is below $\xi^{\ast}$, the buyers become discerning enough that they buy only from sellers who have $\hat{G}$ signals. The condition $\alpha(r-c)>\beta(1-c)$ ensures that good sellers are more likely to be assigned $\hat{G}$ signals and that this advantage is large enough for them to be better off than bad sellers on average, allowing them to grow the share until they reach frequency $\xi^{\ast}$ (see Methods for further details).

\section{Results}

\subsubsection*{Long-Run Equilibrium of the Market}

We focus on results when $\alpha>\beta$. Although signals are labeled $\hat{G}$ and $\hat{B}$ for convenience, they are arbitrary binary signals, and the results for cases when $\alpha<\beta$ directly follow by flipping the role of signals (or redefining the probabilities as $\alpha'\triangleq 1-\alpha$ and $\beta'\triangleq 1-\beta$). Importantly, if $\alpha(r-c)>\beta(1-c)$ is satisfied, then there is a unique asymptotically stable equilibrium for the evolution of seller types; and both good and bad sellers coexist at the equilibrium. The frequency of good sellers in this equillibrium is given by
\begin{equation*}
\xi^{\ast} = \frac{1-\beta}{r(1-\alpha)+(1-\beta)},
\end{equation*}
which validates our intuition that equillibrium share of good sellers increases when $\alpha$ is higher and $\beta$ is lower, corresponding to more accurate signals (see also Methods).

The unique stable equilibrium arises from the interplay between buyer decision-making and seller behavioral adaptation. Consider first a situation where most active sellers in the population are good. Such a population must not be stable because, if only a minute portion of sellers in the market are bad, buyers would purchase from any seller, irrespective of the signal, as the expected payoff is high in such environment. Bad sellers would then enjoy  advantage, gaining greater profit than good sellers. This allows invasion and growth of bad sellers until the share of good sellers is pushed down to the equilibrium value $\xi^{\ast}$. On the other hand, when the share of good sellers is below $\xi^{\ast}$, the buyers become discerning enough that they buy only from sellers who have $\hat{G}$ signals. The condition $\alpha(r-c)>\beta(1-c)$ ensures that good sellers are more likely to be assigned $\hat{G}$ signals and that this advantage is large enough for them to be better off than bad sellers on average, allowing them to grow the share until they reach frequency $\xi^{\ast}$ (see Methods for further details).

\subsection*{Platform Incentives}

\subsubsection*{Platform incentives with costless signaling}

\begin{figure}[H]
\centering
\includegraphics[width=0.52\linewidth]{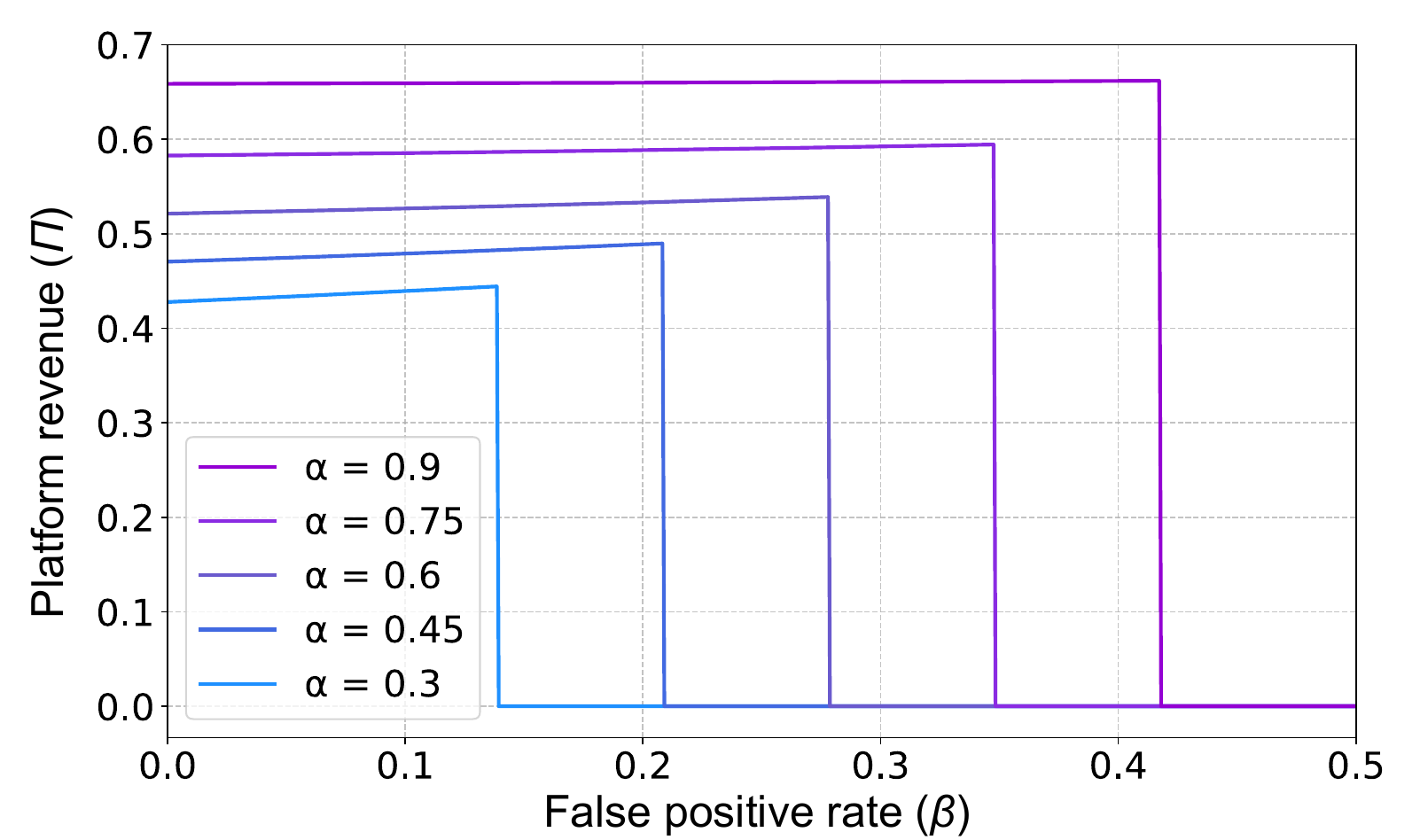}
\caption{\small Platform revenue $\Pi$ as a function of true-positive and false-positive rates, $\alpha$ and $\beta$. Revenue rises with the true-positive rate because more accurate recognition of good sellers increases their equilibrium share and boosts successful transactions. Revenue also increases with the false-positive rate, since bad sellers make more sales when they are mislabeled as good. However, this effect is bounded: beyond an upper threshold  $\bar \beta$, buyer trust collapses and revenue falls to zero. The figure illustrates both the incentive for the platform to promote transactions by blurring distinctions between seller types, and the constraint that too much dishonesty undermines the stability of the market. Parameters: $r = 0.85, c = 0.72$ with $C(\alpha,\beta)\equiv 0$.}
\label{fig:pi_select_alpha}
\end{figure}

Crucially, under the assumption $\alpha>\beta$, the platform's revenue is increasing with respect to \emph{both} $\alpha$ and $\beta$ when there is no cost to modifying accuracy, $C(\alpha,\beta)\equiv0$ (Figure~\ref{fig:pi_select_alpha}). It is straightforward to see why profit increases with the true-positive rate $\alpha$. The equilibrium share of good sellers increases with $\alpha$, and there is a better chance for good sellers to successfully make sales than bad sellers (since good sellers are more likely to be assigned a good signal when $\alpha>\beta$), leading to a higher revenue for the platform. The incentives of the platform concerning the false-positive rate $\beta$ are more nuanced. Although a higher $\beta$ reduces the equilibrium share of good sellers, it also boosts the sales of the bad sellers, resulting in a net increase in platform revenue (see Methods for details). 

The most obvious incentive for the platform is to achieve at least some positive revenue. To achieve this, the platform must ensure the existence of an internal stable equilibrium of seller types, or otherwise no transactions will occur at all. (Recall that the state in which all sellers are good--though it would yield positive revenue---cannot constitute a stable equilibrium: buyers come to believe that all sellers are good irrespective of the signal, which gives bad sellers an advantage and enables their invasion.) And so the platform will always choose signaling probabilities to ensure $\alpha(r-c)> \beta(1-c)$. This renders the platform's incentives largely aligned with market integrity, at least when it is cost-free to provide accurate information (i.e., when $C(\alpha,\beta)\equiv0$). 

Specifically, in this case, it is optimal for the platform to be perfectly honest about good sellers, meaning that the platform maximizes its profit by setting $\alpha= 1$. Although it will be perfectly honest about good sellers, the platform may nonetheless set a value of $\beta$ greater than zero, meaning it will sometimes assign signal $\hat{G}$ to bad sellers. Nevertheless, the constraint $\alpha(r-c)> \beta(1-c)$, which ensures a non-zero rate of sales,  enforces an upper bound on the false-positive rate $\beta$ that is compatible with the platform's incentive:
\begin{equation*}
\beta<\bar{\beta}\triangleq\alpha\cdot\frac{r-c}{1-c}
\end{equation*}
which reduces to $\beta<(r-c)/(1-c)$ when $\alpha=1$. This means there is a limit to the extent to which the platform can be dishonest. It is easy to see that upper bound $\bar{\beta}$ is increasing in $r$ and decreasing in $c$---which means the platform can remain profitable while increasing the rate of false-positive signals when the benefit of successful transactions is large or the commission fee is small. When the platform is perfectly accurate about good sellers, $\alpha=1$, the platform will be indifferent to any false-positive rate $\beta$ satisfying $\beta<\bar{\beta}$ (see the derivative $\partial \Pi/\partial \beta$ in Methods). Thus, the platform can maximize its profit for any value of false-positive rate $\beta$ less than this upper bound. Notably, the range of incentive-compatible $\beta$ includes $\beta= 0$, in which case the reputation system is perfectly accurate and the platform still maximizes its profit.

\subsubsection*{Platform incentives with costly signaling}

The platform's incentives are fundamentally different in the more realistic scenario when accurate signaling is costly. In particular, the result that a platform can maximize profit while being perfectly honest does not generally hold when it is costly to maintain an accurate reputation system. 

To explore the effects of costly accuracy, we consider the following plausible form of cost function:
\begin{equation*}
\begin{split}
C(\alpha,\beta)
=\left\{
\begin{array}{ll}
0 & \textrm{if } (\alpha,\beta)\in S(\alpha_{0},\beta_{0})  \\
\kappa\cdot \frac{d(\alpha,\beta,S)^{p}}{\left(\alpha\beta(1-\alpha)(1-\beta)\right)^{q}} & \textrm{otherwise}
\end{array}
\right.
\end{split}
\end{equation*}
where $\kappa>0$, $q\leq 1\leq p$ and $ S(\alpha_{0},\beta_{0})=\textrm{conv}\{(0,0),(1,1),(\alpha_{0},\beta_{0}),(1-\alpha_{0},1-\beta_{0})\}$ is the convex hull of the accuracy pairs obtainable for free: the natural accuracy \((\alpha_{0},\beta_{0})\), the accuracy obtained by flipping the sign \((1-\alpha_{0},1-\beta_{0})\), and those achievable by random signaling along the line \(\alpha = \beta\). Because combining two free strategies via randomization requires no additional resources, it is natural to deem the entire convex hull of these points cost‑free. 

To make this cost function concrete, we may interpret the displayed signals of the reputation system as generated by a classifier $h$ that outputs $h(G)=\hat{G}$ with probability $\alpha$ and $h(B)=\hat{G}$ with probability $\beta$, which  correspond to the classifier's true positive and false positive rates, respectively. Given two classifiers $h_{1}$ and $h_{2}$ with accuracies $(\alpha_{1},\beta_{1})$ and $(\alpha_{2},\beta_{2})$, a classifier of accuracy $\gamma(\alpha_{1},\beta_{1})+(1-\gamma)(\alpha_{2},\beta_{2})$ can be obtained by forming $h_{\gamma}$ that returns the output of $h_{1}$ with probability $\gamma$ and that of $h_{2}$ with probability $1-\gamma$.

\begin{figure}[H]
\centering
\includegraphics[width=0.4\linewidth]{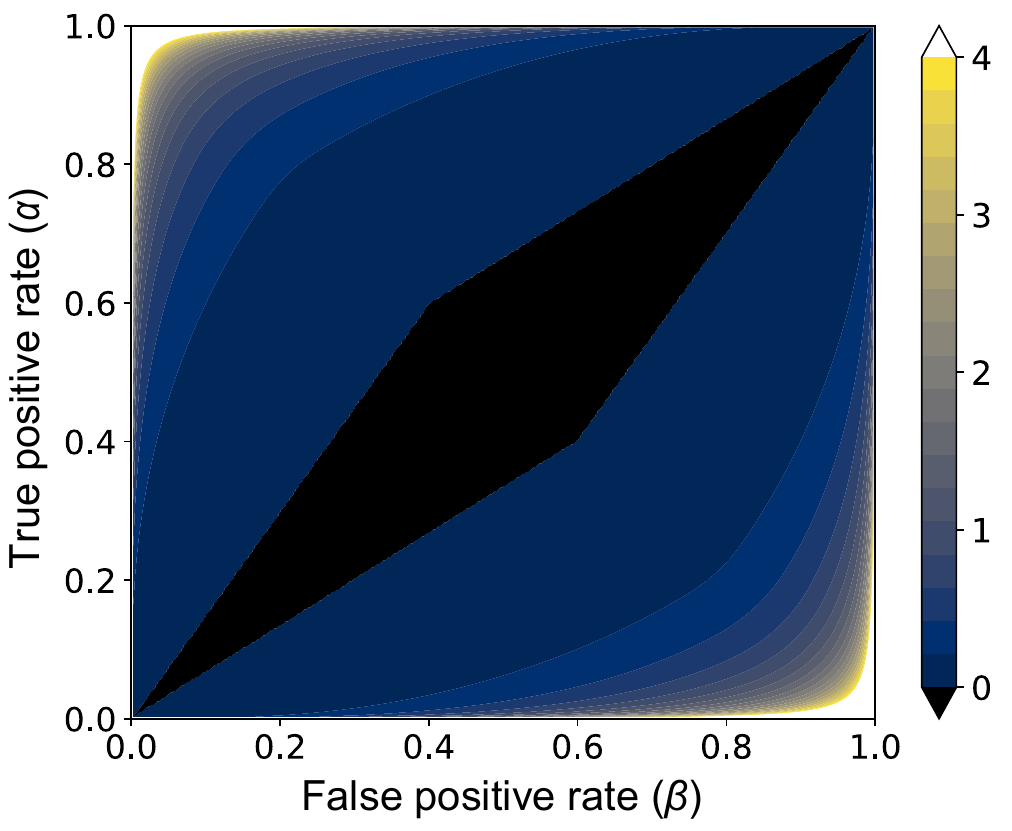}
\caption{\small The cost $C(\alpha,\beta)$ to a platform for maintaining a reputation system with true-positive rate $\alpha$ and false-positive rate $\beta$. The cost of the reputation system is zero in the convex hull of freely achievable pairs, which includes the ``natural" values $(\alpha_0,\beta_0)$ as well as any pairs with $\alpha=\beta$. Outside the convex hull, the cost grows as the accuracy increases. Parameters: $\alpha_{0}=0.6,\beta_{0}=0.4,\kappa=0.5,p=2,q=0.5$.}
\label{fig:platform_cost}
\end{figure}

Outside the cost-free convex hull, we have stipulated that the cost grows with the distance $d$ from the convex hull, defined as:
\begin{equation*}
d(\alpha,\beta,S)=\underset{P\in\partial S}{\min} \lVert (\alpha,\beta)-P \rVert
\end{equation*}
where $\partial S$ denotes the boundary of $S$ and $\lVert \cdot \rVert$ is the Euclidean norm. The denominator forces the cost to approach infinity as $\alpha$ approaches 0 or 1 
and likewise for $\beta$, reflecting the practical difficulty in achieving perfectly accurate signals about either good or bad sellers. This cost function remains constant along each ROC (Receiver Operating Characteristic) level curve (Figure~\ref{fig:platform_cost}), but it escalates as accuracy increases, approaching infinity in the limit of perfect classifier performance.

\begin{figure}
\centering
\includegraphics[width=0.4\linewidth]{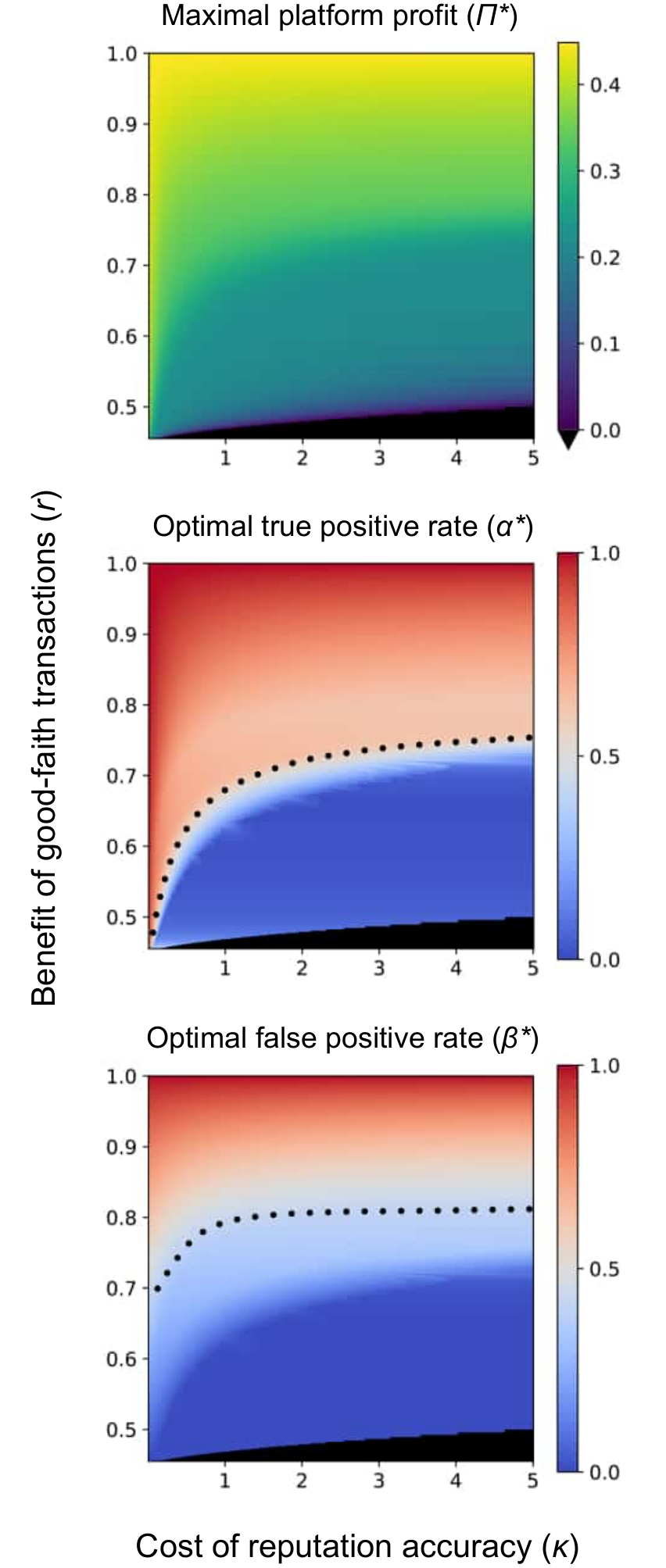}
\caption{\small 
Maximal platform profit (top) and optimal true-positive  (middle) and false-positive (bottom) rates. Dotted lines mark the baseline ``natural'' accuracy levels, ($\alpha_{0}, \beta_{0}$). Platform profit increases when good-faith transactions generate more value (higher $r$) but falls when the cost of improving accuracy ($\kappa$) is high. The platform’s optimal strategy generally involves investing to raise both $\alpha$ and $\beta$ above their natural levels. Notably, the optimal false-positive rate $\beta^{\ast}$ generally increases with $r$ and it can even exceed its baseline value $\beta_{0}$, meaning that the platform has a net benefit from paying to artificially inflate the ratings of bad sellers. When $r$ is low and $\kappa$ is high, the cost of improving signals dominates, and the platform refrains from investing in accuracy altogether, resulting in zero profit (black regions). Parameter: $c=0.45$, $p=2$, $q=0.5$ $\alpha_{0}=0.6$, $\beta_{0}=0.4$.}
\label{fig:pi_alpha_beta_rk}
\end{figure}

Even when providing accurate signals is costly, the constraint $\alpha(r-c)>\beta(1-c)$ must still hold for the platform to earn any positive revenue, so the same upper bound $\bar{\beta}$ on $\beta$ from the cost-free scenario remains in effect. Nevertheless, the values of $\alpha$ and $\beta$ that maximize the platform's profit are \emph{both} above the natural accuracy,  $(\alpha_{0},\beta_{0})$, over a wide range of values of exogenous parameters, as shown in Figure~\ref{fig:pi_alpha_beta_rk}. Notably, the profit-maximizing value of $\beta$ is also above its baseline value, $\beta_{0}$, which means there is an incentive for platform-side rating inflation: the platform maximizes net profit by paying a cost to make it more difficult to distinguish between good and bad sellers. 

Because it is costly to achieve a high true-positive rate $\alpha$, and $\alpha=1$ is practically infeasible (except when $\beta=1$, which conflicts with the platform's incentive constraints due to the upper bound $\bar{\beta}$), the platform tends to obtain the highest profit at some intermediate value of $\alpha$. When $\alpha<1$, since the platform earns higher revenue (i.e., the total amount of commission collected) with higher $\beta$, the platform-optimal value of $\beta$ will generally exceed $0$ and can even exceed $\beta_{0}$, depending on the actual form of the cost function. The platform's optimal value of $\beta$ tends to increase when buyers and good sellers gain more profit from sales (parameter $r$), because $r$ raises the upper bound $\bar{\beta}$. The optimal value of $\beta$ also tends to increase when accuracy cost rate $\kappa$ becomes smaller, because the increase in revenue from moving $\beta$ above $\beta_{0}$ is more likely to outweigh its cost.

\subsubsection*{Variable commission fee}

\begin{figure}
\centering
\includegraphics[width=0.4\linewidth]{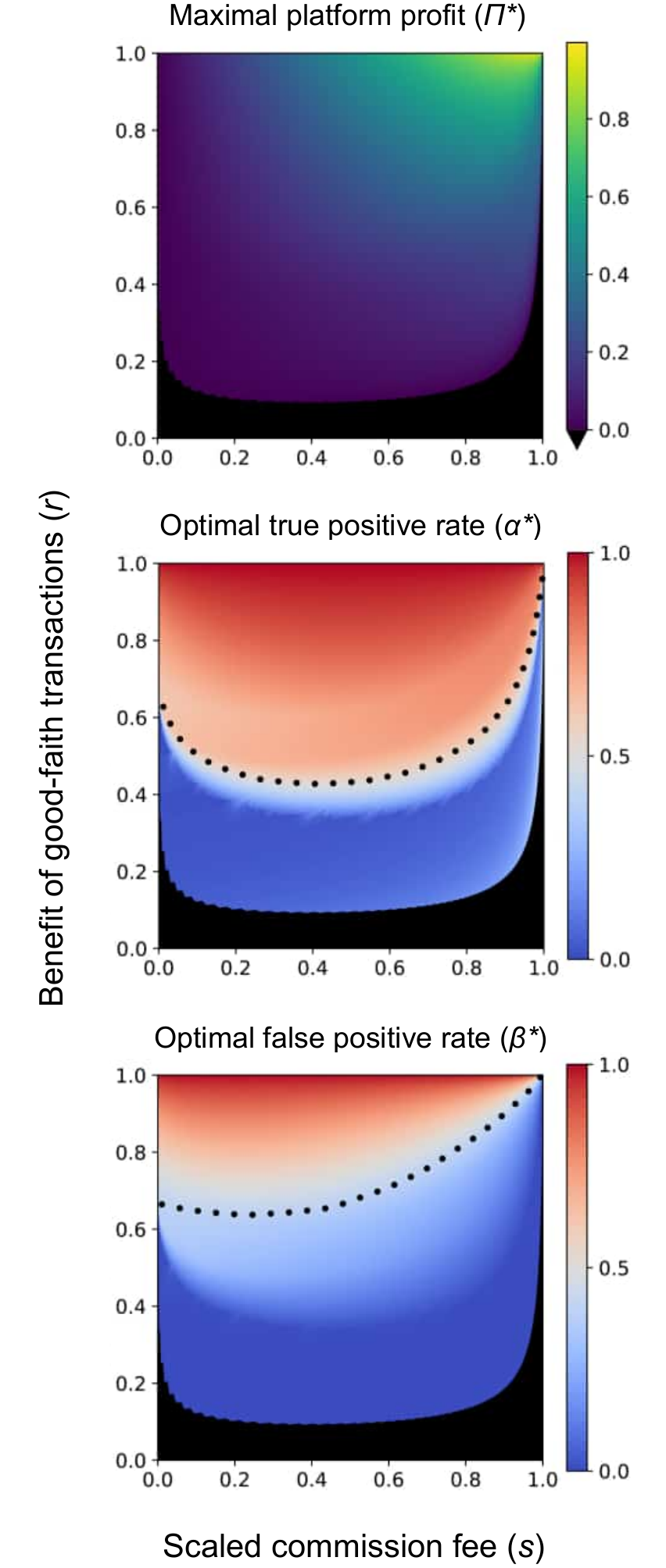}
\caption{\small Maximal platform profit (top) and optimal true-positive  (middle) and false-positive (bottom) rates, in response to the scaled commission fee $s \triangleq c/r$. Dotted lines indicate the natural accuracy $\alpha_{0}$ and $\beta_{0}$. When the platform can adjust the commission fee $c$, for a given value of $r$, it tends to obtain higher profit with a higher commission $s$, provided it does not approach the extreme case $s=1$. In the region with high $s$, the optimal value of $\beta$ is much lower than the optimal $\alpha$, indicating the incentive to maintain a somewhat accurate reputation system. The region where the platform incurs no cost and gains no profit is indicated in black. Parameter: $\kappa=0.2$, $p=2$, $q=0.5$, $\alpha_{0}=0.6$, $\beta_{0}=0.4$.}
\label{fig:pi_alpha_beta_rs}
\end{figure}

So far we have treated the commission $c$ paid from the sellers to the platform as an exogenous parameter. We may alternatively think of this as an endogenous variable set by the platform. We define $s \triangleq c/r$ to be the ratio of commission fee to the benefit of good-faith transactions; and we analyze the platform-optimal value of the scaled commission fee, together with the corresponding optimal parameters of the reputation system $\alpha$ and $\beta$ (Figures~\ref{fig:pi_alpha_beta_rs} and~S3). Note that $\alpha(r-c)> \beta(1-c)$ is clearly violated when $s\geq 1$, so $s$ must lie in $[0,1)$ to be compatible with the incentive of the platform.

For a fixed value of $r$ and $\kappa$, the platform tends to maximize its revenue with a high value of scaled commission $s$, typically above $0.8$ in the numerical example we have considered (Figure~S3
). Since $\bar{\beta}$ is decreasing in $c$, the optimal $\beta$ often falls below $\beta_{0}$, due to the lower value of $\bar{\beta}$ when $s$ is high (and thus $c$ is high relative to $r$). Consequently, for a wide range of $r$, the optimal $\alpha$ exceeds $\alpha_{0}$ and the optimal $\beta$ is less than $\beta_{0}$ (Figure~S3
). This indicates that the incentive for rating inflation is mitigated (in fact, there is pressure to increase accuracy) when the platform can freely set the commission fee it obtains from sales.

\subsection*{Social utility for market participants}

So far we have focused on what is optimal for the platform, which generally has the ability the modify the nature of its reputation signaling system, perhaps at a cost. 
But the platform's reputation system also has important consequences for the welfare of the buyers and sellers who participate in the marketplace. To study this, for different values of signaling fidelity $\alpha$ and $\beta$, we will compute the resulting average payoff of the buyers post-transaction, $U^{b}$, as well as the average payoff of the sellers $U^{s}$ defined as:
\begin{equation*}
\begin{split}
U^{b}&\triangleq \left(\alpha+\frac{1}{2}(1-\alpha)\right)\xi^{\ast}r+\left(\beta+\frac{1}{2}(1-\beta)\right)(1-\xi^{\ast})(-1)\\
U^{s}&\triangleq\xi^{\ast}\pi_{G}+(1-\xi^{\ast})\pi_{B}.
\end{split}
\end{equation*}
Note that we consider utility to marketplace participants after the frequencies of good versus bad seller types have reached equilibrium in response to the platform's policies ($c$, $\alpha$, $\beta$).

\begin{figure}[H]
\centering
\includegraphics[width=0.38 \linewidth]{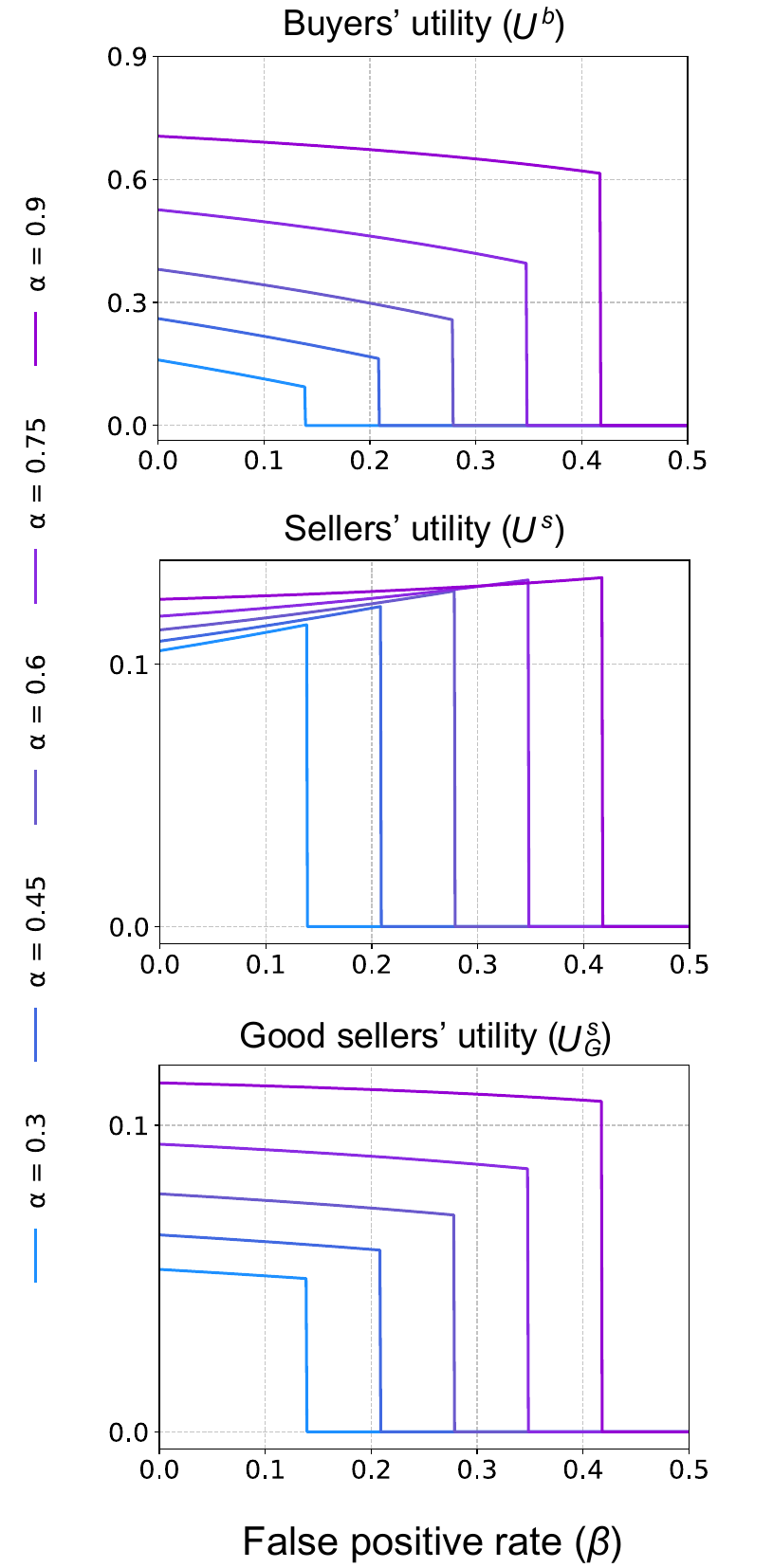}
\caption{\small Social utility produced by a reputation system. Buyer utility $U^b$ increases with the true-positive rate $\alpha$ and decreases with the false-positive rate $\beta$. The same pattern holds when focusing on the utility of good sellers $U^s_G$. 
By contrast, the average utility of sellers as a whole $U^s$ can move in the opposite direction, because untrustworthy sellers profit from higher $\beta$, while trustworthy sellers profit from higher $\alpha$.  Parameters: $r=0.85, c=0.72$.}
\label{fig:ub_us_usg}
\end{figure}

We find that the welfare of the buyers and the sellers as a whole can be conflicted (Figure \ref{fig:ub_us_usg}). Buyers benefit more when the information is more accurate, as $U^{b}$ is increasing in $\alpha$ and decreasing in $\beta$. On the other hand, sellers can benefit from inaccurate signals. Specifically, $U^{s}$ is increasing with respect to $\beta$, whereas it is typically increasing with respect to $\alpha$---see Methods for details on these partial derivatives. 

When considering social utility, it may be reasonable to ignore the utility provided to bad sellers, whose behavior is morally objectionable. If we consider only the utility provided to good sellers, which is captured by
\begin{equation*}
U^{s}_{G}\triangleq\xi^{\ast}\pi_{G},
\end{equation*}
we find that it is always increasing in $\alpha$ and decreasing in $\beta$, in alignment with buyers' utility (Figure \ref{fig:ub_us_usg}). Thus, as might be expected, a platform provides higher social benefits to good-faith actors when its reputation system is more accurate. 

These results highlight an intrinsic conflict between the incentives of a platform that broadcasts reputation information and the general welfare of the buyers and faithful sellers who operate within the marketplace provided. We have already seen that, from the platform's perspective, it is often profit-maximizing to artificially reduce the accuracy of reputation signals, especially those concerning bad sellers and especially when good-faith transactions are more mutually beneficial for buyers and sellers (Figure~\ref{fig:pi_alpha_beta_rk}). However, this platform behavior tends to diminish the marketplace's overall utility for both buyers and faithful sellers (Figure~ \ref{fig:ub_us_usg}).

When the platform is able to optimize the commission paid by the seller, $c$, along with the signal probabilities $\alpha$ and $\beta$, the incentives of the platform largely align with the buyers, but not with the sellers. As a result, the utilities of buyers and sellers, including those of bad sellers, can conflict with each other when commission fee $c$ is under platform control (Figure~S4 
top versus middle). When the platform chooses the values of $\alpha$ and $\beta$ to maximize its profit, sellers' utility declines as $c$ increases, whereas buyers can benefit from a more accurate reputation system supported by higher fees (Figure S4
).

\section{Discussion}

Institutions that broadcast reputations in a marketplace confront two intertwined moral hazards. The first is the classic hazard of trust between buyers and sellers: buyers cannot directly observe seller types, and sellers face incentives to defect after payment. The second arises at the institutional level: the platform itself may compromise accuracy of the reputation system, exaggerating positive signals to encourage transactions that boost its commission revenue. Our model captures both hazards by embedding a noisy reputation system within a platform-mediated trust game. This framework reveals how the platform’s choices about signal accuracy shape the composition of sellers, the willingness of buyers to engage, and ultimately the alignment---or misalignment---between institutional profit and social welfare.

Our model extends a large body of theoretical work on the evolution of trust \cite{Lim2024Trust,hu2021,liu2023,masuda2012,liu2025}, social reputations \cite{radzvilavicius2021,hu2021,liu2023,liepanis2024,kessinger2024institutionspublicjudgmentestablished,liepanis2024}, cooperation fostered by institutions \cite{radzvilavicius2021,kessinger2024institutionspublicjudgmentestablished,chibaokabe2024}, and reputations in online marketplaces \cite{kwark2014,kwark2017,dellarocas2006,mayzlin2006,duke2016,jiang2020,li2017,shi2023}. Much of this prior work has examined how institutions influence cooperation by assigning rewards or broadcasting reputations, while models of online platforms have analyzed how reviews and disclosure policies shape market outcomes. By bringing these strands together, our framework captures both sides of the problem: the adaptive behavior of buyers and sellers in a trust game, and the profit-driven incentives of the platform itself. This combination allows us to study the feedback loop in which platform policies shape seller conduct and buyer trust, which in turn feed back into the platform’s revenues and strategic choices.

Our model of institutional reputations and participant behavior is also related to the literature on Bayesian persuasion \cite{kamenica2011bayesian,kamenica2019}. Although that literature typically deals with different contexts, it suggests that a platform can design a reputation system to steer Bayesian-rational buyers toward actions that benefit the platform. Studies in Bayesian persuasion usually assume that the prior belief (corresponding to the proportion of good sellers in our model) is set exogenously, whereas our model incorporates economic incentive–driven market dynamics that endogenize the prior belief so that it responds to the parameters of the reputation system.

Under our model, for a given accuracy of the reputation system (represented by the true- and false-positive rates $\alpha$ and $\beta$) that satisfies a certain conditions, there is a unique internal equilibrium in which good and bad sellers stably coexist. This is in fact the only equilibrium in which the platform earns a positive revenue. The platform's revenue increases when  good and bad sellers are more indistinguishable, which indicates an incentive for the platform to allow rating inflation. This is because bad sellers, even if harmful to buyers, nonetheless generate commission revenue for the platform, provided buyers continue to purchase. However, there exists a limit to how inaccurate the reputation system can be while still being profitable. If the reputation system becomes too inaccurate, there will be fewer good sellers at equilibrium, eventually causing buyers to lose trust altogether. This tension---between wanting inflated ratings to boost overall transactions and needing sufficient accuracy to keep good sellers and trusting buyers in the market---creates a boundary on how far the platform can afford to skew the ratings in favor of bad sellers. When a higher level of accuracy of the reputation system is accompanied by correspondingly higher cost, as it would likely be the case in the real world, the platform may achieve the highest net profit by paying a cost to artificially reduce the accuracy of its reputation system.

Commission fees paid by sellers to the platform further complicate the platform’s strategy. Higher commission fees naturally increase the platform’s per-transaction profit but also amplify the need to preserve an advantage for good sellers, who have lower margins than bad sellers. Hence, to set a higher commission fee, the platform has to maintain a more accurate reputation system to keep good sellers viable. From the perspective of social welfare, buyers benefit from a more accurate system while sellers are negatively impacted by higher commission fees. As a result, the platform ends up increasing fees and offering high utility to buyers but low utility to sellers. In practice, however, there may be limits to the platform's choice of commission fees, particularly when it must compete not only with other similar platforms but also with alternative selling channels, such as traditional physical retail stores and direct-to-consumer sales, to attract market participants.

Our analysis provides one potential explanation for rating inflation \cite{nosko2015,zervas2020airbnb,filippas2022} from the perspective of platform incentives, complementing previous studies that have examined other explanations \cite{li2008,dellarocas2008,bolton2013,nosko2015,fradkin2021,filippas2022,muchnik2013,wang2018,brown2006,mayzlin2014,xu2015}. In doing so, we adopt several simplifying assumptions. Notably, reputations are treated as binary and memoryless: sellers are assigned either signal $\hat{G}$ or signal $\hat{B}$ in each time period, with no persistence across time. By contrast, online marketplaces typically implement persistent, multi-level reputation systems (e.g., cumulative 5-star ratings), which allow reputations to accumulate over time and influence future interactions. In addition, we have not directly modeled the process of artificial manipulation of ratings by sellers, and the possibility that a buyer may engage in repeated purchases from the same seller.

Under these simplifications, our analysis provides a comprehensible framework to understand the incentives of the platform and its effects on the marketplace. This account
reveals an intricate interplay between the platform behavior and resulting market dynamics---including the incentive for the platform to artificially inflate ratings. The extent of such distortion is constrained by the reaction of market participants, and it is also disciplined by the platform’s commission-setting incentives. While our study has focused on a simple setting, where we can derive intuition and robust qualitative predictions, future research may explore more complicated and realistic scenarios, including how competition among institutions and heterogeneity within the population of agents influence the overall market ecology.

\section{Methods}

\subsection{Market Dynamics}

The configuration of the reputation system impacts the behavior of market participants. We first analyze the decision-making of buyers. Given $\alpha,\beta$ and $\xi$, the probability of the randomly matched seller's type being $G$ conditioned on observing signal $\hat{G}$, or $P(G|\hat{G})$, can be computed as:
\begin{equation*}
P(G|\hat{G})=\frac{P(\hat{G}|G)P(G)}{P(\hat{G}|G)P(G)+P(\hat{G}|B)(1-P(G))}=\frac{\alpha \xi}{(\alpha-\beta)\xi+\beta}.
\end{equation*}
Similarly, $P(G|\hat{B})=(1-\alpha)\xi/[(\beta-\alpha)\xi+(1-\beta)]$. Thus, the expected payoff of a buyer who visited the platform is:
\begin{equation*}
E[R|\hat{G}]=rP(G|\hat{G})+(-1)(1-P(G|\hat{G}))=\frac{(1+r)\alpha \xi}{(\alpha-\beta) \xi +\beta}-1
\end{equation*}
or
\begin{equation*}
E[R|\hat{B}]=rP(G|\hat{B})+(-1)(1-P(G|\hat{B}))=\frac{(1+r)(1-\alpha)\xi}{(\beta-\alpha)\xi+(1-\beta)}-1,
\end{equation*}
depending on the reputation signal observed by the buyer. 
Here $R$ denotes the random variable representing buyers' payoff.
The conditions for a purchase to be made with probability 1 can be written, respectively, as:
\begin{equation*}
\begin{split}
&E[R|\hat{G}]>0\Leftrightarrow \xi>\xi^{\hat{G}}\triangleq\frac{\beta}{r\alpha+\beta}\\
&E[R|\hat{B}]>0\Leftrightarrow \xi> \xi^{\hat{B}}\triangleq\frac{1-\beta}{r(1-\alpha)+(1-\beta)}.
\end{split}
\end{equation*}
The buyer purchases with probability $1/2$ if $\xi=\xi^{\hat{G}}$ or $\xi=\xi^{\hat{B}}$, depending on the observed signal. 

This decision-making pattern, in turn, determines the payoff of sellers. Given the reputation signals and corresponding purchase probabilities, good and bad sellers' expected payoffs are:
\begin{equation*}
\begin{split}
\pi_{G}^{\hat{G}}&=
\left\{
\begin{array}{ll}
 r-c & \textrm{if } \xi>\xi^{\hat{G}}\\
 \frac{1}{2}(r-c) &\textrm{if }\xi=\xi^{\hat{G}}\\
 0 & \textrm{otherwise}
\end{array}
\right.\\
\pi_{G}^{\hat{B}}&=
\left\{
\begin{array}{ll}
r-c & \textrm{if }\xi> \xi^{\hat{B}}\\
\frac{1}{2}(r-c) &\textrm{if }\xi=\xi^{\hat{B}}\\
0 & \textrm{otherwise}
\end{array}
\right.\\
\pi_{B}^{\hat{G}}&=
\left\{
\begin{array}{ll}
 1-c & \textrm{if } \xi> \xi^{\hat{G}}\\
 \frac{1}{2}(1-c)&\textrm{if }\xi= \xi^{\hat{G}} \\
 0 & \textrm{otherwise}
\end{array}
\right.\\
\pi_{B}^{\hat{B}}&=
\left\{
\begin{array}{ll}
1-c & \textrm{if }\xi> \xi^{\hat{B}}\\
\frac{1}{2}(1-c) & \textrm{if }\xi=\xi^{\hat{B}}\\
0 & \textrm{otherwise}.
\end{array}
\right.\\
\end{split}
\end{equation*}
Since:
\begin{equation*}
\xi^{\hat{G}}-\xi^{\hat{B}}=\frac{r(\beta-\alpha)}{(r\alpha+\beta)(r(1-\alpha)+(1-\beta))},
\end{equation*}
we have $\xi^{\hat{G}}<\xi^{\hat{B}}$ if $\alpha>\beta$ whereas $\xi^{\hat{G}}>\xi^{\hat{B}}$ if $\beta>\alpha$. Therefore, if $\alpha>\beta$:
\begin{equation*}
\begin{split}
\pi_{G}&=\left\{
\begin{array}{ll}
r-c& \textrm{if } \xi > \xi^{\hat{B}} \\
\alpha (r-c)+(1-\alpha)\frac{1}{2}(r-c)=\frac{(1+\alpha)(r-c)}{2}&\textrm{if } \xi=\xi^{\hat{B}}\\
\alpha (r-c) +(1-\alpha)\cdot 0=\alpha (r-c)& \textrm{if } \xi^{\hat{B}}> \xi >\xi^{\hat{G}}\\
\alpha \frac{1}{2}(r-c)+(1-\alpha)\cdot 0=\frac{\alpha (r-c)}{2}&\textrm{if }\xi=\xi^{\hat{G}}\\
0&\textrm{if }\xi^{\hat{G}}>\xi
\end{array}
\right.\\
\pi_{B}&=\left\{
\begin{array}{ll}
1-c& \textrm{if } \xi > \xi^{\hat{B}} \\
\beta\cdot (1-c)+(1-\beta)\frac{1}{2}(1-c)=\frac{(1+\beta)(1-c)}{2}&\textrm{if }\xi=\xi^{\hat{B}}\\
\beta\cdot (1-c)+(1-\beta)\cdot 0=\beta(1-c)& \textrm{if } \xi^{\hat{B}}> \xi >\xi^{\hat{G}}\\
\beta\frac{1}{2}(1-c)+(1-\beta)\cdot 0 =\frac{\beta(1-c)}{2}&\textrm{if }\xi=\xi^{\hat{G}}\\
0&\textrm{if }\xi^{\hat{G}}>\xi.
\end{array}
\right.
\end{split}
\end{equation*}
On the other hand, if $\beta>\alpha$:
\begin{equation*}
\begin{split}
&\pi_{G}=\left\{
\begin{array}{ll}
\alpha (r-c) + (1-\alpha)(r-c)=r-c& \textrm{if } \xi > \xi^{\hat{G}} \\
\alpha\frac{1}{2}(r-c)+ (1-\alpha)(r-c)=(1-\frac{\alpha}{2})(r-c) &\textrm{if }\xi = \xi^{\hat{G}}\\
\alpha\cdot 0 +(1-\alpha)(r-c)=(1-\alpha) (r-c)& \textrm{if } \xi^{\hat{G}}> \xi >\xi^{\hat{B}}\\
\alpha\cdot 0+(1-\alpha)\frac{1}{2}(r-c)=\frac{(1-\alpha)(r-c)}{2}&\textrm{if }\xi=\xi^{\hat{B}}\\
\alpha\cdot 0+(1-\alpha)\cdot 0=0&\textrm{if }\xi^{\hat{B}}>\xi
\end{array}
\right.\\
&\pi_{B}=\left\{
\begin{array}{ll}
\beta\cdot (1-c)+(1-\beta)\cdot (1-c)=1-c& \textrm{if } \xi > \xi^{\hat{G}} \\
\beta\frac{1}{2}(1-c)+(1-\beta)\cdot (1-c)= (1-\frac{\beta}{2})(1-c)&\textrm{if }\xi=\xi^{\hat{G}}\\
\beta\cdot 0+(1-\beta)\cdot (1-c)=(1-\beta)(1-c)& \textrm{if } \xi^{\hat{G}}> \xi >\xi^{\hat{B}}\\
\beta\cdot 0+(1-\beta)\frac{1}{2}(1-c)=\frac{(1-\beta)(1-c)}{2}&\textrm{if }\xi=\xi^{\hat{B}}\\
\beta\cdot 0 +(1-\beta)\cdot 0=0&\textrm{if }\xi^{\hat{B}}>\xi.
\end{array}
\right.
\end{split}
\end{equation*}

We focus our analysis on the case $\alpha>\beta$ as the results for $\beta>\alpha$ follow by flipping the roles of the signals. Since $x_{B}\equiv 1-x_{G}-x_{I}$, the behavior of the replicator equations (Equation~\ref{eq:replicator}) can be described by a system of two equations. Keeping in mind that $\pi_{I}=0$, the system of equations reduces to:
\begin{equation*}
\begin{split}
\dot{x_{G}}&=x_{G}\left(\pi_{G}-\sum_{i}x_{i}\pi_{i}\right)\\
&=x_{G}[\pi_{G}-(x_{G}\pi_{G}+(1-x_{G}-x_{I})\pi_{B})]\\
&=x_{G}[(1-x_{G})(\pi_{G}-\pi_{B})+x_{I}\pi_{B}]\\
\dot{x_{I}}&=x_{I}\left(\pi_{I}-\sum_{i}x_{i}\pi_{i}\right)\\
&=-x_{I}(x_{G}\pi_{G}+(1-x_{G}-x_{I})\pi_{B}).
\end{split}
\end{equation*}
Technically, $P(G|\hat{B})$ is undefined under the conditions $\alpha=1$ and $\xi=1$. This, in turn, results in $\pi_{B}$ being undefined on the measure-zero subset of the state space where $\xi=1$ is satisfied when $\alpha=1$. To avoid the system of ODE being undefined as a result, we let $\dot{x_{G}}\triangleq x_{G}(\pi_{G}-(x_{G}\pi_{G}+x_{I}\pi_{I})) = x_{G}(1-x_{G})\pi_{G}$ and $\dot{x_{I}}\triangleq x_{I}(\pi_{I}-(x_{G}\pi_{G}+x_{I}\pi_{I})) = -x_{I}x_{G}\pi_{G}$ in this case. These definitions match the formal expressions of the replicator equations for any finite $\pi_{B}$ when $\xi=1$ (and, thus, $x_{B}=0$) and preserve the dynamics. Similarly, $\pi_{G}$ is undefined when $\beta=1$ and $\xi=0$ are both true, so we let $\dot{x_{G}}\triangleq x_{G}(\pi_{G}-(x_{G}\pi_{G}+x_{I}\pi_{I})) = 0$ and $\dot{x_{I}}\triangleq x_{I}(\pi_{I}-(x_{G}\pi_{G}+x_{I}\pi_{I})) = -x_{I}x_{B}\pi_{B}$ on the corresponding subset of the state space as well.

Notice that, for the existence of a non-trivial internal stable equilibrium---in which any transaction actually takes place--- requires $r-c>0$. If $r-c<0$, good sellers are driven out of the market, leaving only bad and inactive sellers, meaning $\xi=0$. And if $r=c$, the system does not have a stable equilibrium as the good and inactive sellers would always have the same payoff. If $r-c>0$, the payoff of both good and bad sellers are strictly greater than that of inactive sellers provided $\alpha,\beta>0$. Thus, the system invariably moves towards the set of states in which only active sellers exist ($x_{I}=0$), and the growth rate of inactive sellers fixates at $\dot{x_{I}}=0$ once it reaches the set, effectively reducing the system to a single-variable differential equation $\dot{x_{G}}=x_{G}[(1-x_{G})(\pi_{G}-\pi_{B})]$. Since $x_{G}=\xi$ in this case, $\dot{x_{G}}=\dot{\xi}=\xi(1-\xi)(\pi_{G}-\pi_{B})$. 

Technically, the replicator equation in our model has points of discontinuity at $\xi^{\hat{G}}$ and $\xi^{\hat{B}}$, which may prevent it from having a classical solution. Thus, we consider solutions in Filippov sense \cite{filippov1988differential,cortes2008}. Specifically, denoting the reduced ODE by $\dot{\xi}=\Xi(\xi(t))$, we consider the reduced differential inclusion $\dot{\xi}\in F[\Xi](\xi)$ where $F[\Xi](\xi)$ is the Filippov set-valued map given by:
\begin{equation*}
\begin{split}
&F[\Xi](\xi)=\left\{
\begin{array}{ll}
\xi(1-\xi)(r-1) &\textrm{if }\xi>\xi^{\hat{B}}  \\
\left[\xi(1-\xi)(r-1),\, \xi(1-\xi)(\alpha(r-c)-\beta(1-c))\right] &\textrm{if }\xi=\xi^{\hat{B}}  \\
\xi(1-\xi)(\alpha(r-c)-\beta(1-c))&\textrm{if }\xi^{\hat{B}}>\xi>\xi^{\hat{G}}\\
\left[0,\,\xi(1-\xi)(\alpha(r-c)-\beta(1-c))\right]&\textrm{if }\xi=\xi^{\hat{G}}\\
0&\textrm{if }\xi^{\hat{G}}>\xi.
\end{array}
\right.
\end{split}
\end{equation*}

There is an internal equilibrium at $\xi=\xi^{\hat{B}}$ if both $\beta<\alpha$ and $\alpha(r-c)>\beta(1-c)$ hold. This is the only asymptotically stable equilibrium, and, in particular, $\xi=1$ (all good seller) is always unstable. The dynamics of the system can be intuitively understood by plotting $\Delta\triangleq \pi_{G}-\pi_{B}$ (Figure S5
). Since $\xi(1-\xi)$ is always positive in the interior, the share of good sellers increases if $\Delta>0$ and decreases if $\Delta<0$ relative to bad sellers. We know that $\Delta<0$ when $\xi>\xi^{\hat{B}}$, thereby decreasing $\xi$, and $\Delta=0$ when $\xi<\xi^{\hat{G}}$, so that the dynamics is driven solely by random fluctuation (neutral drift). The overall dynamics is then determined by the sign of $\Delta$ in $\xi^{\hat{G}}<\xi<\xi^{\hat{B}}$. If $\Delta>0$, $\xi$ is pushed up or down until it reaches $\xi^{\hat{B}}$ and stay there. On the other hand, if $\Delta<0$, it gets pushed down to $\xi^{\hat{G}}$. Neutral drift will only allow $\xi$ to move within the area at or below $\xi^{\hat{G}}$, and the dynamics eventually end up at $\xi^{\hat{B}}$ in the former case or stay in $[0,\xi^{\hat{G}}]$ in the latter case. No purchases are made below $\xi^{\hat{G}}$, resulting in zero payoff to the platform.

\subsection{Platform Incentives}

We analyze the platform incentive by assuming $\kappa=0$ (and so $C(\alpha,\beta)=0$), so that $\Pi$ remains analytically tractable. Given $\alpha>\beta$, the platform can sustainably gain strictly positive profit only when $\alpha(r-c)>\beta(1-c)$. This leads to the unique stable internal equilibrium at $\xi^{\hat{B}}$, in which $p^{\hat{G}}_{G}=p^{\hat{G}}_{B}=1$ and $p^{\hat{B}}_{G}=p^{\hat{B}}_{B}=1/2$. Therefore,
\begin{equation*}
\begin{split}
\Pi &= c\left(\left[\alpha+\frac{1}{2}(1-\alpha)\right] \xi^{\hat{B}}+\left[\beta+\frac{1}{2}(1-\beta)\right]\left(1-\xi^{\hat{B}}\right)\right)\\
&=\frac{2(1-\alpha) cr+(1-\beta) c[2-(1-\alpha)(1+r)]}{2[(1-\alpha) r+(1-\beta)]}
\end{split}
\end{equation*}
if $\alpha(r-c)>\beta(1-c)$ and otherwise $\Pi=0$.

On the other hand, when $\beta>\alpha$,
\begin{equation*}
\begin{split}
\Pi&= 
c\left(\left[\frac{1}{2}\alpha+(1-\alpha)\right]\xi^{\hat{G}}+\left[\frac{1}{2}\beta+(1-\beta)\right]\left(1-\xi^{\hat{G}}\right)\right)\\
&=\frac{2\alpha cr+\beta c(2-\alpha(1+r))}{2(\alpha r+\beta)}
\end{split}
\end{equation*}
if $(1-\alpha)(r-c)>(1-\beta)(1-c)$ is also satisfied and otherwise $\Pi=0$.

Notice that the first expression is essentially equivalent to the second (flipping $\alpha$ to $1-\alpha$ and same for $\beta$). Because of this, the relation between the platform profit and reputation system accuracy is symmetric (specifically, the profit is decreasing with respect to both $\alpha$ and $\beta$) to the $\alpha>\beta$ case. We proceed with our analysis by assuming $\alpha>\beta$ as the results for $\beta>\alpha$ follow from symmetry). Taking the derivative of $\Pi$ with respect to $\alpha$ and $\beta$ yields:
\begin{equation*}
\begin{split}
\frac{\partial \Pi}{\partial \alpha}&=\frac{(1-\beta)^{2}c(1+r)}{2[(1-\beta)+(1-\alpha)r]^{2}}\geq0\\
\frac{\partial \Pi}{\partial \beta}&=\frac{(1-\alpha)^{2}cr(1+r)}{2[(1-\beta)+(1-\alpha)r]^{2}} \geq 0
\end{split}
\end{equation*}
where the inequalities are strict when $\alpha\neq 1$ and $\beta\neq 1$, respectively. This indicates that the platform is incentivized to increase both $\alpha$ and $\beta$.

\subsection{Social Utility}

Assuming $\alpha>\beta$, utility of sellers and buyers can be written  explicitly as:
\begin{equation*}
\begin{split}
U^{s}&=\xi^{\hat{B}}\pi_{G}+(1-\xi^{\hat{B}})\pi_{B}\\
&=\xi^{\hat{B}}\left[\alpha\pi_{G}^{\hat{G}}+(1-\alpha)\pi_{G}^{\hat{B}}\right]+(1-\xi^{\hat{B}})\left[\beta\pi^{\hat{G}}_{B}+(1-\beta)\pi_{B}^{\hat{B}}\right]\\
&=\frac{1}{2}\left(\xi^{\hat{B}}\left[(1+\alpha)(r-c)-(1+\beta)(1-c)\right]+(1+\beta)(1-c)\right)\\
&=\frac{1}{2}\left(\frac{(1-\beta)\left[(1+\alpha)(r-c)-(1+\beta)(1-c)\right]}{r(1-\alpha)+(1-\beta)}+(1+\beta)(1-c)\right)\\
U^{b}&=\frac{1}{2}\left(\xi^{\hat{B}}\left[(1+\alpha)r+(1+\beta)\right]-(1+\beta)\right)\\
&=\frac{1}{2}\left(\frac{(1-\beta)[(1+\alpha)r+(1+\beta)]}{r(1-\alpha)+(1-\beta)}-(1+\beta)\right).
\end{split}
\end{equation*}

Taking the derivatives:
\begin{equation*}
\begin{split}
\frac{\partial U^{b}}{\partial \alpha}=\frac{(1-\beta)r(1+r)}{[(1-\alpha)r+(1-\beta)]^{2}}\geq 0\\
\frac{\partial U^{b}}{\partial \beta}=-\frac{(1-\alpha)r(1+r)}{[(1-\alpha)r+(1-\beta)]^{2}}\leq 0
\end{split}
\end{equation*}
where the inequalities are strict when $\beta\neq 1$ and $\alpha\neq 1$, respectively. Therefore, buyer utility increases as the reputation system becomes more accurate.

On the other hand, sellers can benefit from inaccurate signals. Specifically, we have: 
\begin{equation*}
\begin{split}
\frac{\partial U^{s}}{\partial \alpha}=&-\frac{(1-\beta)[(1-\beta)c(1+r)-2r(r-\beta)]}{2[r(1-\alpha)+(1-\beta)]^{2}}< 0 \\
&\textrm{ if }\left\{ 
\begin{array}{ll}
 \frac{2r^{2}}{1+r}<c\textrm{; or} \\
 c\leq\frac{2r^{2}}{1+r} \textrm{ and }\frac{c(1+r)-2r^{2}}{c(1+r)-2r}<\alpha,\beta
\end{array}
\right.\\
\frac{\partial U^{s}}{\partial \beta}=&\frac{(1-\alpha)r[2(1-\alpha r)-(1-\alpha)c(1+r)]}{2[r(1-\alpha)+(1-\beta)]^{2}} \geq 0.
\end{split}
\end{equation*}

Specifically focusing on good sellers:
\begin{equation*}
\begin{split}
U^{s}_{G}&=\xi^{\hat{B}}\left[\alpha\pi_{G}^{\hat{G}}+(1-\alpha)\pi_{G}^{\hat{B}}\right]\\
&=\frac{(1-\beta)(1+\alpha)(r-c)}{2[r(1-\alpha)+(1-\beta)]}.
\end{split}
\end{equation*}

Taking derivatives:
\begin{equation*}
\begin{split}
\frac{\partial U^{s}_{G}}{\partial \alpha}&=\frac{(1-\beta)(1-\beta+2r)(r-c)}{2[r(1-\alpha)+(1-\beta)]^{2}}\geq 0\\
\frac{\partial U^{s}_{G}}{\partial \beta}&=-\frac{(1-\alpha)(1+\beta)r(r-c)}{2[r(1-\alpha)+(1-\beta)]^{2}}\leq 0
\end{split}
\end{equation*}
where the inequalities are strict when $\beta\neq 1$ and $\alpha\neq 1$, respectively. Thus, the utility of good sellers aligns with the buyers.

\printbibliography

\newpage

\renewcommand{\thefigure}{S\arabic{figure}}
\renewcommand{\thetable}{S\arabic{table}}
\renewcommand{\thesubsection}{S\arabic{subsection}}
\setcounter{figure}{0} 
\setcounter{table}{0}
\setcounter{subsection}{0}

\section*{Supplementary Information}

\subsection*{Derivation of the system of ODEs from a stochastic process}

Our replicator equation (Eq.~1 in main text) can be seen as an approximation of a stochastic process in a large but finite population on every point of continuity. In this section, we describe the stochastic process and the derivation of the system of ODEs from that process following a standard method \cite{traulsen2006}.

Let $N$ be the number of sellers in the population and $N_{G}$ and $N_{B}$ be the number of good and bad sellers, respectively. At the beginning of each time period, good sellers are assigned $\hat{G}$ ($\hat{B}$) with probability $\alpha$ ($1-\alpha$), and bad sellers are assigned $\hat{G}$ ($\hat{B}$) with probability $\beta$ ($1-\beta$). Each time period lasts for some duration $t$ where buyers arrive at rate $\lambda$, following a Poisson process. At the end of each time period, one seller is randomly picked as a focal individual $j$ who considers updating its strategy. Another seller, say $j'$, is randomly picked with whom the focal individual compares their payoffs. The focal individual $j$ switches its type to that of $j'$ with a probability given by the Fermi function:
\begin{equation*}
\phi(\bar{\pi}_{j}(n_{j})-\bar{\pi}_{j'}(n_{j'}))=\frac{1}{1+\exp[\sigma (\bar{\pi}_{j}(n_{j})-\bar{\pi}_{j'}(n_{j'}))]}
\label{eq:fermi}
\end{equation*}
where $\sigma$ is a parameter that determines the strength of selection and $\bar{\pi}_{j},\bar{\pi}_{j'}$ are the payoffs of the focal and the other sellers, respectively, averaged across all the sales each individual made in that time period. Variables $n_{j},n_{j'}$ are the number of sales made by each individual in that time period. When no sales were made, the average is treated as $0$. By Taylor expansion, we obtain the approximation for the Fermi function for small $\sigma$:
\begin{equation}
\begin{split}
\phi(x)&=\frac{1}{1+\exp(\sigma x)}\\
&= \left.\frac{1}{1+\exp(\sigma x)}\right|_{\sigma=0}+\sigma\left.\left(\frac{d}{d\sigma}\left[\frac{1}{1+\exp(\sigma x)}\right]\right)\right|_{\sigma=0}+\mathcal{O}(\sigma^{2})\\
&\approx \frac{1}{2}-\frac{\sigma x}{4}.
\end{split}
\end{equation}
We denote this approximation for convenience by $\widetilde{\phi}(x)=1/2-\sigma x/4$.

For any fixed one-off payoff $\pi$, since $\bar{\pi}(n)=\left(\frac{1}{n}\cdot n\pi\right)=\pi$ for any $n>0$, the average payoff is simply $\pi$ with probability $\sum_{n=1}^{\infty}\left[\left(\frac{\lambda t}{N}\right)^{n}/n!\right]e^{-\frac{\lambda t}{N}}=1-e^{-\Lambda}$ and $0$ with probability $e^{-\Lambda}$ where $\Lambda=\lambda t/N$. Thus, even though there are infinitely many combination for the possible number of sales made by two individuals, $j$ and $j'$, there are only four ($2\times 2$) possible combinations of payoffs, the probabilities of which are easy to calculate. For any pair of one-off payoffs $\pi,\pi'$ and the corresponding number of sales $n,n'$, keeping in mind that $\bar{\pi}(n)=\bar{\pi}(m)$ for any $n,m>0$ as we just observed,
\begin{equation*}
\begin{split}
&\sum_{n=0}^{\infty}\frac{\Lambda^{n}}{n!}e^{-\Lambda}\sum_{n'=0}^{\infty}\frac{\Lambda^{n'}}{n'!}e^{-\Lambda}\widetilde{\phi}(\bar{\pi}(n)-\bar{\pi}'(n'))\\
&=(1-e^{-\Lambda})^{2}\widetilde{\phi}(\bar{\pi}(1)-\bar{\pi}'(1))+(1-e^{-\Lambda})e^{-\Lambda}\widetilde{\phi}(\bar{\pi}(1)-\bar{\pi}'(0))\\
&\quad +e^{-\Lambda}(1-e^{-\Lambda})\widetilde{\phi}(\bar{\pi}(0)-\bar{\pi}'(1))+(e^{-\Lambda})^{2}\widetilde{\phi}(\bar{\pi}(0)-\bar{\pi}'(0))\\
&=\frac{1}{2}-\frac{\sigma}{4}[(1-e^{-\Lambda})^{2}(\pi-\pi')+(1-e^{-\Lambda})e^{-\Lambda}(\pi-0)+e^{-\Lambda}(1-e^{-\Lambda})(0-\pi')+(e^{-\Lambda})^{2}\cdot0]\\
&=\frac{1}{2}-(1-e^{-\Lambda})\frac{\sigma(\pi-\pi')}{4}.
\end{split}
\end{equation*}
Thus, we simply rescale $\sigma$ by dividing it by $1-e^{-\Lambda}$ to account for the variability in the payoff arising from the number of buyers visiting each seller.

The stochastic system can be characterized by transition probabilities $T^{+}_{G},T^{-}_{G},T^{+}_{B},T^{-}_{B}$ where $T^{+}_{G}$ represents the probability that the number of good sellers increase by one, and so on. The switching probability given by the Fermi function (Eq.~\ref{eq:fermi}) varies depending on the true type and the signal assigned to the randomly chosen sellers. The transition probabilities can be computed from the probability of occurrences of each configuration of the Fermi function. Taking $T^{+}_{G}$ as an example, replacing $\phi$ with $\widetilde{\phi}$, the transition probability can be computed and approximated as:

\begin{equation*}
\begin{split}
T^{+}_{G}&\approx\sum_{k=0}^{N_{B}}\binom{N_{B}}{k}\beta^{k}(1-\beta)^{N_{B}-k}\sum_{l=0}^{N_{G}}\binom{N_{G}}{l}\alpha^{l}(1-\alpha)^{N_{G}-l}\\
&\quad \left[\frac{k}{N}\frac{l}{N}\widetilde{\phi}(\pi_{B}^{\hat{G}}-\pi_{G}^{\hat{G}})+\frac{N_{B}-k}{N}\frac{l}{N}\widetilde{\phi}(\pi_{B}^{\hat{B}}-\pi_{G}^{\hat{G}})\right.\\
&\quad+\frac{k}{N}\frac{N_{G}-l}{N}\widetilde{\phi}(\pi_{B}^{\hat{G}}-\pi_{G}^{\hat{B}})+\frac{N_{B}-k}{N}\frac{N_{G}-l}{N}\widetilde{\phi}(\pi_{B}^{\hat{B}}-\pi_{G}^{\hat{B}})\\
&\left.\quad +\frac{N-N_{B}-N_{G}}{N}\frac{l}{N}\widetilde{\phi}(\pi_{I}-\pi^{\hat{G}}_{G})+\frac{N-N_{B}-N_{G}}{N}\frac{N_{G}-l}{N}\widetilde{\phi}(\pi_{I}-\pi^{\hat{B}}_{G})\right]\\
&=\frac{N_{B}}{N}\frac{N_{G}}{N}\sum_{k=0}^{N_{B}}\binom{N_{B}}{k}\beta^{k}(1-\beta)^{N_{B}-k}\sum_{l=0}^{N_{G}}\binom{N_{G}}{l}\alpha^{l}(1-\alpha)^{N_{G}-l}\\
&\quad \left[\frac{k}{N_{B}}\frac{l}{N_{G}}\widetilde{\phi}(\pi_{B}^{\hat{G}}-\pi_{G}^{\hat{G}})+\frac{N_{B}-k}{N_{B}}\frac{l}{N_{G}}\widetilde{\phi}(\pi_{B}^{\hat{B}}-\pi_{G}^{\hat{G}})\right.\\
&\quad \left.+\frac{k}{N_{B}}\frac{N_{G}-l}{N_{G}}\widetilde{\phi}(\pi_{B}^{\hat{G}}-\pi_{G}^{\hat{B}})+\frac{N_{B}-k}{N_{B}}\frac{N_{G}-l}{N_{G}}\widetilde{\phi}(\pi_{B}^{\hat{B}}-\pi_{G}^{\hat{B}})\right]\\
&\quad +\frac{N-N_{B}-N_{G}}{N}\frac{N_{G}}{N}\sum_{l=0}^{N_{G}}\binom{N_{G}}{l}\alpha^{l}(1-\alpha)^{N_{G}-l}\left[\frac{l}{N_{G}}\widetilde{\phi}(\pi_{I}-\pi_{G}^{\hat{G}})+\frac{N_{G}-l}{N_{G}}\widetilde{\phi}(\pi_{I}-\pi_{G}^{\hat{B}})\right]\\
&=\frac{N_{B}}{N}\frac{N_{G}}{N}\left[\widetilde{\phi}(\pi_{B}^{\hat{G}}-\pi_{G}^{\hat{G}})\sum_{k=0}^{N_{B}} \sum_{l=0}^{N_{G}} \frac{k}{N_{B}} \frac{l}{N_{G}} \binom{N_{B}}{k} \binom{N_{G}}{l} \beta^{k} (1-\beta)^{N_{B}-k} \alpha^{l} (1-\alpha)^{N_{G}-l}\right.\\
&\quad+\widetilde{\phi}(\pi_{B}^{\hat{B}}-\pi_{G}^{\hat{G}})\sum_{k=0}^{N_{B}} \sum_{l=0}^{N_{G}} \frac{N_{B}-k}{N_{B}} \frac{l}{N_{G}} \binom{N_{B}}{k} \binom{N_{G}}{l} \beta^{k} (1-\beta)^{N_{B}-k} \alpha^{l} (1-\alpha)^{N_{G}-l}\\
&\quad+\widetilde{\phi}(\pi_{B}^{\hat{G}}-\pi_{G}^{\hat{B}})\sum_{k=0}^{N_{B}} \sum_{l=0}^{N_{G}} \frac{k}{N_{B}} \frac{N_{G}-l}{N_{G}} \binom{N_{B}}{k} \binom{N_{G}}{l} \beta^{k} (1-\beta)^{N_{B}-k} \alpha^{l} (1-\alpha)^{N_{G}-l}\\
&\left.\quad+\widetilde{\phi}(\pi_{B}^{\hat{B}}-\pi_{G}^{\hat{B}})\sum_{k=0}^{N_{B}} \sum_{l=0}^{N_{G}} \frac{N_{B}-k}{N_{B}} \frac{N_{G}-l}{N_{G}} \binom{N_{B}}{k} \binom{N_{G}}{l} \beta^{k} (1-\beta)^{N_{B}-k} \alpha^{l} (1-\alpha)^{N_{G}-l}\right]\\
&\quad +\frac{N-N_{B}-N_{G}}{N}\frac{N_{G}}{N}\left[\widetilde{\phi}(\pi_{I}-\pi_{G}^{\hat{G}})\sum_{l=0}^{N_{G}}\frac{l}{N_{G}}\binom{N_{G}}{l}\alpha^{l}(1-\alpha)^{N_{G}-l}\right.\\
&\quad\left. +\widetilde{\phi}(\pi_{I}-\pi_{G}^{\hat{B}})\sum_{l=0}^{N_{G}}\frac{N_{G}-l}{N_{G}}\binom{N_{G}}{l}\alpha^{l}(1-\alpha)^{N_{G}-l}\right].
\end{split}
\end{equation*}
Notice that, since $E[L]=N_{G}\alpha$ and $E[K]=N_{B}\beta$ where $L\sim \textrm{Bin}(N_{G},\alpha)$ and $K\sim \textrm{Bin}(N_{B},\beta)$:
\begin{equation*}
\begin{split}
&\sum_{k=0}^{N_{B}} \sum_{l=0}^{N_{G}} \frac{k}{N_{B}} \frac{l}{N_{G}} \binom{N_{B}}{k} \binom{N_{G}}{l} \beta^{k} (1-\beta)^{N_{B}-k} \alpha^{l} (1-\alpha)^{N_{G}-l}\\
&=\sum_{k=0}^{N_{B}}\frac{k}{N_{B}}\binom{N_{B}}{k}\beta^{k}(1-\beta)^{N_{B}-k}\times \sum_{l=0}^{N_{G}}\frac{l}{N_{G}}\binom{N_{G}}{l}\alpha^{l}(1-\alpha)^{N_{G}-l} \\
&= \beta \cdot \alpha.
\end{split}
\end{equation*}
Applying similar computations to other terms as well, we obtain the approximation:
\begin{equation*}
\begin{split}
T^{+}_{G}&\approx x_{B}x_{G}(\beta\alpha\widetilde{\phi}(\pi_{B}^{\hat{G}}-\pi_{G}^{\hat{G}})+(1-\beta)\alpha(\pi_{B}^{\hat{B}}-\pi_{G}^{\hat{G}})\\
&\quad+\beta(1-\alpha)\widetilde{\phi}(\pi_{B}^{\hat{G}}-\pi_{G}^{\hat{B}})+(1-\beta)(1-\alpha)\widetilde{\phi}(\pi_{B}^{\hat{B}}-\pi_{G}^{\hat{B}})) \\
&\quad +x_{I}x_{G}(\alpha\widetilde{\phi}(\pi_{I}-\pi_{G}^{\hat{G}}) +(1-\alpha)\widetilde{\phi}(\pi_{I}-\pi_{G}^{\hat{B}})).
\end{split}
\end{equation*}
Writing out $\widetilde{\phi}$ and simplifying,
\begin{equation*}
\begin{split}
T^{+}_{G}&\approx x_{B}x_{G}\left[\frac{1}{2}-\beta\alpha\left(\frac{\sigma(\pi_{B}^{\hat{G}}-\pi_{G}^{\hat{G}})}{4}\right)-(1-\beta)\alpha\left(\frac{\sigma(\pi_{B}^{\hat{B}}-\pi_{G}^{\hat{G}})}{4}\right)\right.\\
&\quad\left.-\beta(1-\alpha)\left(\frac{\sigma(\pi_{B}^{\hat{G}}-\pi_{G}^{\hat{B}})}{4}\right)\quad-(1-\beta)(1-\alpha)\left(\frac{\sigma(\pi_{B}^{\hat{B}}-\pi_{G}^{\hat{B}})}{4}\right) \right]\\
&\quad+x_{I}x_{G}\left[\frac{1}{2}-\alpha\left(\frac{\sigma(\pi_{I}-\pi_{G}^{\hat{G}})}{4}\right)-(1-\alpha)\left(\frac{\sigma(\pi_{I}-\pi_{G}^{\hat{B}})}{4}\right)\right]\\
&=\frac{x_{B}x_{G}+x_{I}x_{G}}{2}\\
&\quad +\frac{\sigma x_{B}x_{G}}{4}\left[\pi_{G}^{\hat{B}}-\pi_{B}^{\hat{B}}+\alpha(\pi_{G}^{\hat{G}}-\pi_{G}^{\hat{B}})+\beta(\pi^{\hat{B}}_{B}-\pi_{B}^{\hat{G}})\right]-\frac{\sigma x_{I}x_{G}}{4}\left[\pi_{I}-(\alpha\pi^{\hat{G}}_{G}+(1-\alpha)\pi^{\hat{B}}_{G})\right].
\end{split}
\end{equation*}
Similarly,
\begin{equation*}
\begin{split}
T^{-}_{G}&\approx\sum_{l=0}^{N_{G}}\binom{N_{G}}{l}\alpha^{l}(1-\alpha)^{N_{G}-l}\sum_{k=0}^{N_{B}}\binom{N_{B}}{k}\beta^{k}(1-\beta)^{N_{B}-k}\\
&\quad\left[\frac{l}{N}\frac{k}{N}\widetilde{\phi}(\pi_{G}^{\hat{G}}-\pi_{B}^{\hat{G}})+\frac{N_{G}-l}{N}\frac{k}{N}\widetilde{\phi}(\pi_{G}^{\hat{B}}-\pi_{B}^{\hat{G}})\right.\\
&\quad+\frac{l}{N}\frac{N_{B}-k}{N}\widetilde{\phi}(\pi_{G}^{\hat{G}}-\pi_{B}^{\hat{B}})+\frac{N_{G}-l}{N}\frac{N_{B}-k}{N}\widetilde{\phi}(\pi_{G}^{\hat{G}}-\pi_{B}^{\hat{B}})\\
&\left.\quad +\frac{l}{N}\frac{N-N_{G}-N_{B}}{N}\widetilde{\phi}(\pi_{G}^{\hat{G}}-\pi_{I})+\frac{N_{G}-l}{N}\frac{N-N_{G}-N_{B}}{N}\widetilde{\phi}(\pi_{G}^{\hat{B}}-\pi_{I}) \right]\\
&=\frac{N_{G}}{N}\frac{N_{B}}{N}\sum_{l=0}^{N_{G}}\binom{N_{G}}{l}\alpha^{l}(1-\alpha)^{N_{G}-l}\sum_{k=0}^{N_{B}}\binom{N_{B}}{k}\beta^{k}(1-\beta)^{N_{B}-k}\\
&\quad+\left[\frac{l}{N_{G}}\frac{k}{N_{B}}\widetilde{\phi}(\pi_{G}^{\hat{G}}-\pi_{B}^{\hat{G}})+\frac{N_{G}-l}{N_{G}}\frac{k}{N_{B}}\widetilde{\phi}(\pi_{G}^{\hat{B}}-\pi_{B}^{\hat{G}})\right.\\
&\quad+\left.\frac{l}{N_{G}}\frac{N_{B}-k}{N_{B}}\widetilde{\phi}(\pi_{G}^{\hat{G}}-\pi_{B}^{\hat{B}})+\frac{N_{G}-l}{N_{G}}\frac{N_{B}-k}{N_{B}}\widetilde{\phi}(\pi_{G}^{\hat{G}}-\pi_{B}^{\hat{B}})\right]\\
&\quad+\frac{N_{G}}{N}\frac{1-N_{G}-N_{B}}{N}\sum_{l=0}^{N_{G}}\binom{N_{G}}{l}\alpha^{l}(1-\alpha)^{N_{G}-l}\left[\frac{l}{N_{G}}\widetilde{\phi}(\pi_{G}^{\hat{G}}-\pi_{I})+\frac{N_{G}-l}{N_{G}}\widetilde{\phi}(\pi_{G}^{\hat{B}}-\pi_{I})\right]\\
&= x_{G}x_{B}(\alpha\beta\widetilde{\phi}(\pi_{G}^{\hat{G}}-\pi_{B}^{\hat{G}})+(1-\alpha)\beta\widetilde{\phi}(\pi_{G}^{\hat{B}}-\pi_{B}^{\hat{G}})\\
&+\alpha(1-\beta)\widetilde{\phi}(\pi_{G}^{\hat{G}}-\pi_{B}^{\hat{B}})+(1-\alpha)(1-\beta)\widetilde{\phi}(\pi_{G}^{\hat{B}}-\pi_{B}^{\hat{B}}))\\
&\quad +x_{G}x_{I}(\alpha\widetilde{\phi}(\pi_{G}^{\hat{G}}-\pi_{I})+(1-\alpha)\widetilde{\phi}(\pi_{G}^{\hat{B}}-\pi_{I}))\\
&= \frac{x_{G}x_{B}+x_{G}x_{I}}{2}\\
&+\frac{\sigma x_{G}x_{B}}{4}\left[\pi^{\hat{B}}_{B}-\pi^{\hat{B}}_{G}-\alpha(\pi_{G}^{\hat{G}}-\pi^{\hat{B}}_{G})-\beta(\pi_{B}^{\hat{B}}-\pi^{\hat{G}}_{B})\right]+\frac{\sigma x_{I}x_{G}}{4}\left[\pi_{I}-(\alpha\pi^{\hat{G}}_{G}+(1-\alpha)\pi^{\hat{B}}_{G})\right].
\end{split}
\end{equation*}

Denote the probability that the system is in state $\vec{x}=(x_{G},x_{B})$ at time $\tau$ by $P^{\tau}(x_{G},x_{B})$ (note that $x_{I}$ is automatically specified by the relation $x_{G}+x_{B}+x_{I}=1$). We, then, have:

\begin{equation*}
\begin{split}
P^{\tau+1}(\vec{x})-P^{\tau}(\vec{x})&=P^{\tau}(x_{G}-N^{-1},x_{B})T^{+}_{G}(x_{G}-N^{-1},x_{B})+P^{\tau}(x_{G},x_{B}-N^{-1})T^{+}_{B}(x_{G},x_{B}-N^{-1})\\
&\quad+P^{\tau}(x_{G}+N^{-1},x_{B})T^{-}_{G}(x_{G}+N^{-1},x_{B})+P^{\tau}(x_{G},x_{B}+N^{-1})T^{-}_{B}(x_{G},x_{B}+N^{-1})\\
&\quad -\sum_{i}P^{\tau}(\vec{x})T^{+}_{i}(\vec{x})-\sum_{i}P^{\tau}(\vec{x})T^{-}_{i}(\vec{x}).
\end{split}
\end{equation*}
We define $\rho(\vec{x},t) \triangleq NP^{\tau}(\vec{x})$. Rewriting the above expression in terms of $\rho$'s and applying Taylor expansion, we obtain
\begin{equation*}
\begin{split}
\frac{d\rho(\vec{x},t)}{dt}=&-\sum_{i}\frac{\partial}{\partial x_{i}}\left[\left(T^{+}_{i}(\vec{x})-T^{-}_{i}(\vec{x})\right)\rho(\vec{x},t)\right]+\sum_{i}\frac{1}{2}\frac{\partial^{2}}{\partial x_{i}^{2}}\left[\frac{T^{+}_{i}(\vec{x})+T^{-}_{i}(\vec{x})}{N}\rho(\vec{x},t)\right]
\end{split}
\end{equation*}
which has an equivalent Langevin equation
$dX=A(x,t)dt+B(x,t)dW_{t}$
where $W$ is now a two-dimensional Wiener process and 
\begin{equation*}
\begin{split}
A&=
\begin{pmatrix}
T^{+}_{G}(\vec{x})-T^{-}_{G}(\vec{x})\\
T^{+}_{B}(\vec{x})-T^{-}_{B}(\vec{x})
\end{pmatrix}
\\
B &= 
\begin{pmatrix}
\sqrt{\frac{T^{+}_{G}(\vec{x}) + T^{-}_{G}(\vec{x})}{N}} & 0 \\
0 & \sqrt{\frac{T^{+}_{B}(\vec{x}) + T^{-}_{B}(\vec{x})}{N}} 
\end{pmatrix}.
\end{split}
\end{equation*}
Taking the limit of large $N$ the noise terms become negligible, which leads to the system of ODEs:
\begin{equation*}
\dot{x}_{j}=T^{+}_{j}(\vec{x})-T^{-}_{j}(\vec{x}).
\end{equation*}

\begin{figure}[H]
\centering
\includegraphics[width=0.9\linewidth]{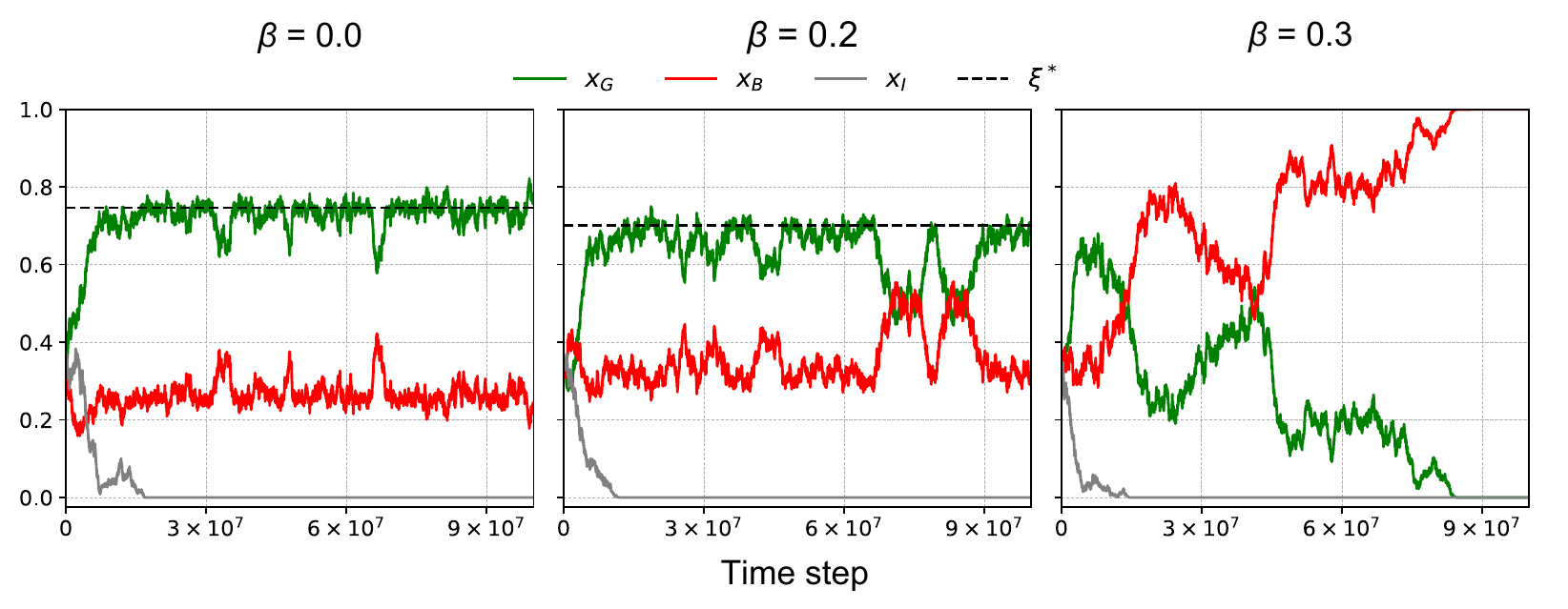}
\caption{\small Simulation of the discrete stochastic model. When $\beta = 0$ and $\beta = 0.2$, the share of inactive sellers drops to zero, and the share of good sellers stabilizes around the equilibrium level $\xi^{\ast}$.
When $\beta = 0.3 > \bar{\beta}$, the internal equilibrium does not exist in the analytical ODE, and the share of good sellers among active sellers is predicted to stay below $\xi^{\hat{G}}$; as in this particular simulation run, good sellers typically go extinct under the stochastic model. Parameters: $r=0.85,c=0.72,\alpha=0.6,N=\Lambda=10,000$.}
\label{fig:simulation}
\end{figure}

\begin{figure}[H]
\centering
\includegraphics[width=0.6\linewidth]{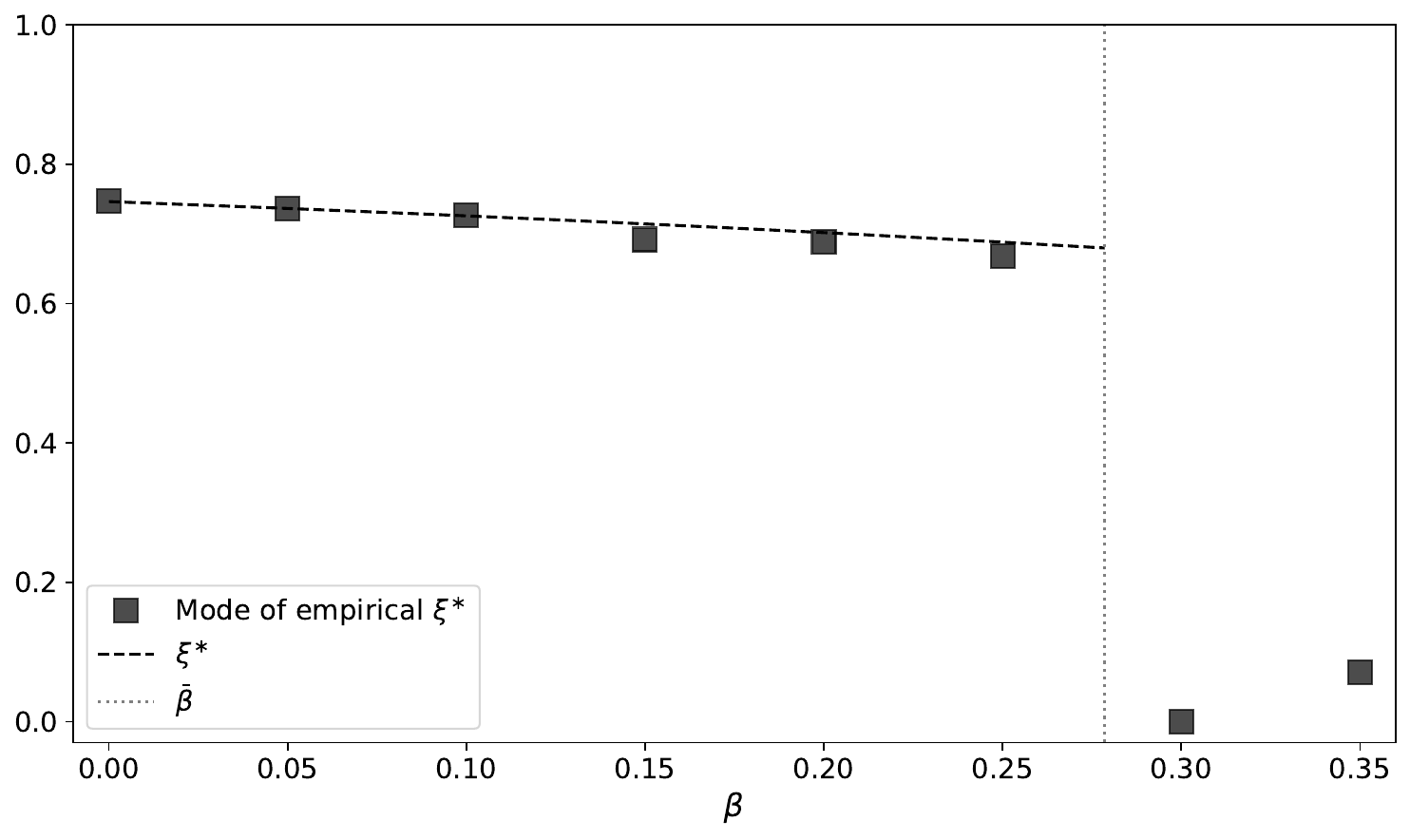}
\caption{\small Summary of simulation results. The plot shows the modes of the empirical $\xi^{\ast}$ for different values of $\beta$, calculated from the last quarter of the simulation time steps during which the system has stabilized in its quasi-stationary regime. The mode corresponds to the most frequently visited state in this period and aligns closely with the analytical model’s prediction, which closely aligns with $\xi^{\ast}$ under the analytical model. Parameters: $r=0.85,c=0.72,\alpha=0.6,N=\Lambda=10,000$.}
\label{fig:empirical_summary}
\end{figure}

Finally, from the results above, we can compute the difference as:
\begin{equation*}
\begin{split}
T^{+}_{G}-T^{-}_{G}&\approx\frac{\sigma}{2}\left[x_{G}x_{B}\left(\pi_{G}^{\hat{B}}-\pi_{B}^{\hat{B}}+\alpha(\pi_{G}^{\hat{G}}-\pi_{G}^{\hat{B}})+\beta(\pi^{\hat{B}}_{B}-\pi_{B}^{\hat{G}})\right)\right.\\
&\left.\quad-x_{G}x_{I}\left(\pi_{I}-(\alpha\pi^{\hat{G}}_{G}+(1-\alpha)\pi^{\hat{B}}_{G})\right)\right]\\
&=\frac{\sigma}{2}x_{G}\left[x_{B}\left(\pi_{G}^{\hat{B}}-\pi_{B}^{\hat{B}}+\alpha(\pi_{G}^{\hat{G}}-\pi_{G}^{\hat{B}})+\beta(\pi^{\hat{B}}_{B}-\pi_{B}^{\hat{G}})\right)-x_{I}\left(\pi_{I}-(\alpha\pi^{\hat{G}}_{G}+(1-\alpha)\pi^{\hat{B}}_{G})\right)\right.\\
&\left.\quad+x_{G}\left(\alpha\pi^{\hat{G}}_{G}+(1-\alpha)\pi_{G}^{\hat{B}}\right)-x_{G}\left(\alpha\pi^{\hat{G}}_{G}+(1-\alpha)\pi_{G}^{\hat{B}}\right)\right]\\
&=\frac{\sigma}{2}x_{G}\left[(x_{G}+x_{B}+x_{I})\left(\alpha\pi_{G}^{\hat{G}}+(1-\alpha)\pi_{G}^{\hat{B}}\right)\right.\\
&\left.\quad-x_{G}\left(\alpha\pi^{\hat{G}}_{G}+(1-\alpha)\pi^{\hat{B}}_{G}\right)-x_{B}\left(\beta x_{B}^{\hat{G}}+(1-\beta)\pi_{B}^{\hat{B}}\right)-x_{I}\pi_{I}\right]\\
&=\frac{\sigma}{2}x_{G}\left[\pi_{G}-x_{G}\pi_{G}-x_{B}\pi_{B}-x_{I}\pi_{I}\right].
\end{split}
\end{equation*}
By repeating similar procedures for other types, we obtain the replicator equation (Eq.~1 in main text) with appropriate rescaling. Computer simulation confirms that the quasi-stationary states of the stochastic process align well with the equilibrium of the analytic system of ODEs (Fig.~\ref{fig:simulation} and Fig.~\ref{fig:empirical_summary}). However, note that in the discrete model, the state $\xi$ typically cannot take the exact value $\xi = \xi^{\ast}$ due to the discreteness of the state space. As a result, the system fluctuates between regimes above and below $\xi^{\ast}$. The platform’s revenue then depends on the proportion of time the system spends in each regime, which means that the platform’s incentive under the discrete stochastic model may deviate from the analytical model, depending on parameter values. Accordingly, the results in this paper are to be interpreted in the limit of a large population.

\bigskip

\begin{figure}[H]
\centering
\includegraphics[width=0.5\linewidth]{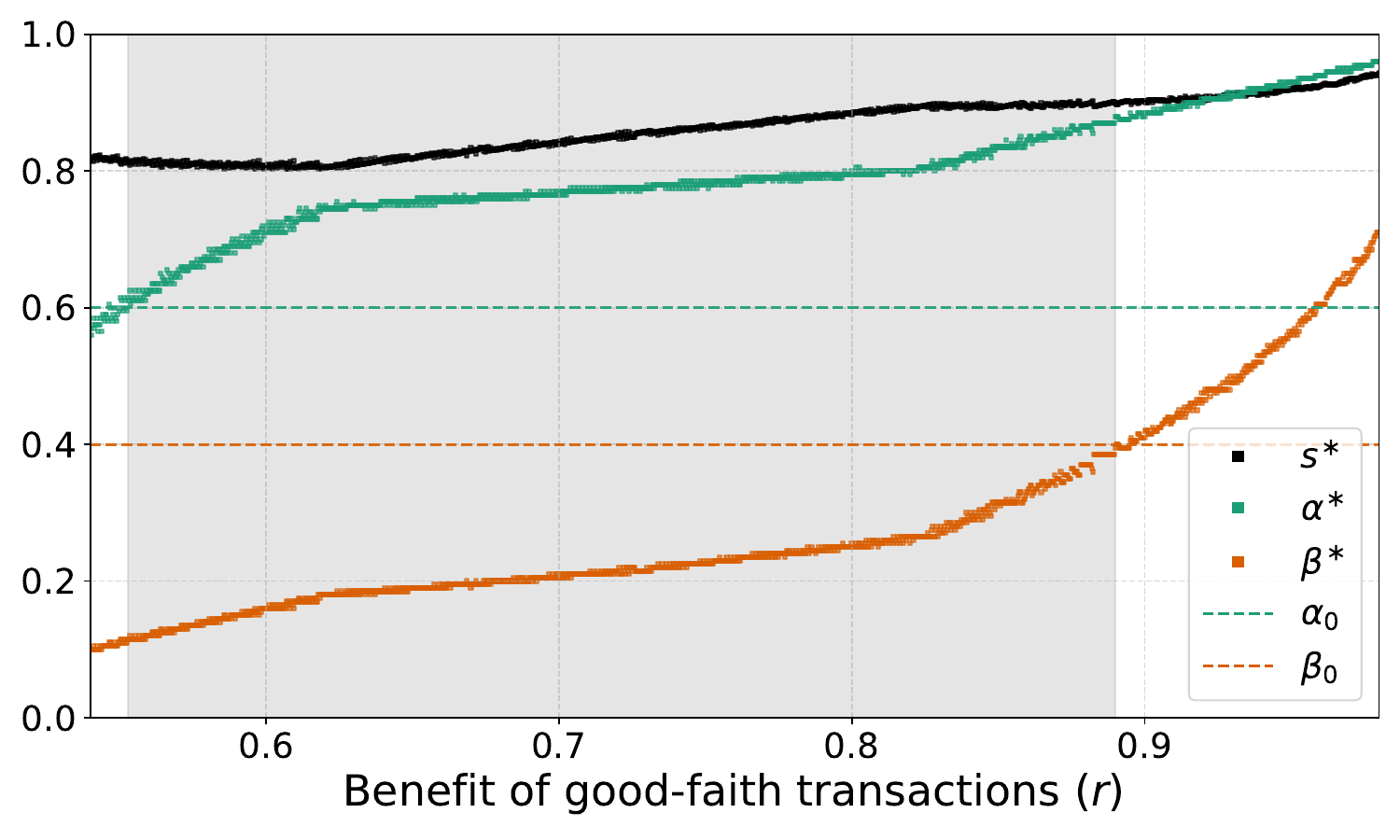}
\caption{\small Optimal values of $s$, $\alpha$, and $\beta$ that maximize platform profit. The underlying information overlaps with Figure~\ref{fig:pi_alpha_beta_rs}, but directly illustrates how the platform’s strategy shifts with commission levels. Across a wide range of $r$, the optimal solution satisfies $\alpha^{\ast} > \alpha_{0}$ and $\beta^{\ast} < \beta_{0}$, showing that the platform’s incentive to inflate ratings is mitigated and even reversed, when it can set its own commission fee. The shaded gray region highlights the parameter space where both inequalities hold, i.e., where the platform will pay to increase signal accuracy for both good and bad sellers. Parameter: $\kappa=0.2$, $p=2$, $q=0.5$, $\alpha_{0}=0.6$, $\beta_{0}=0.4$.}
\label{fig:optimal_s_alpha_beta}
\end{figure}

\clearpage

\begin{figure}[H]
\centering
\includegraphics[width=0.4\linewidth]{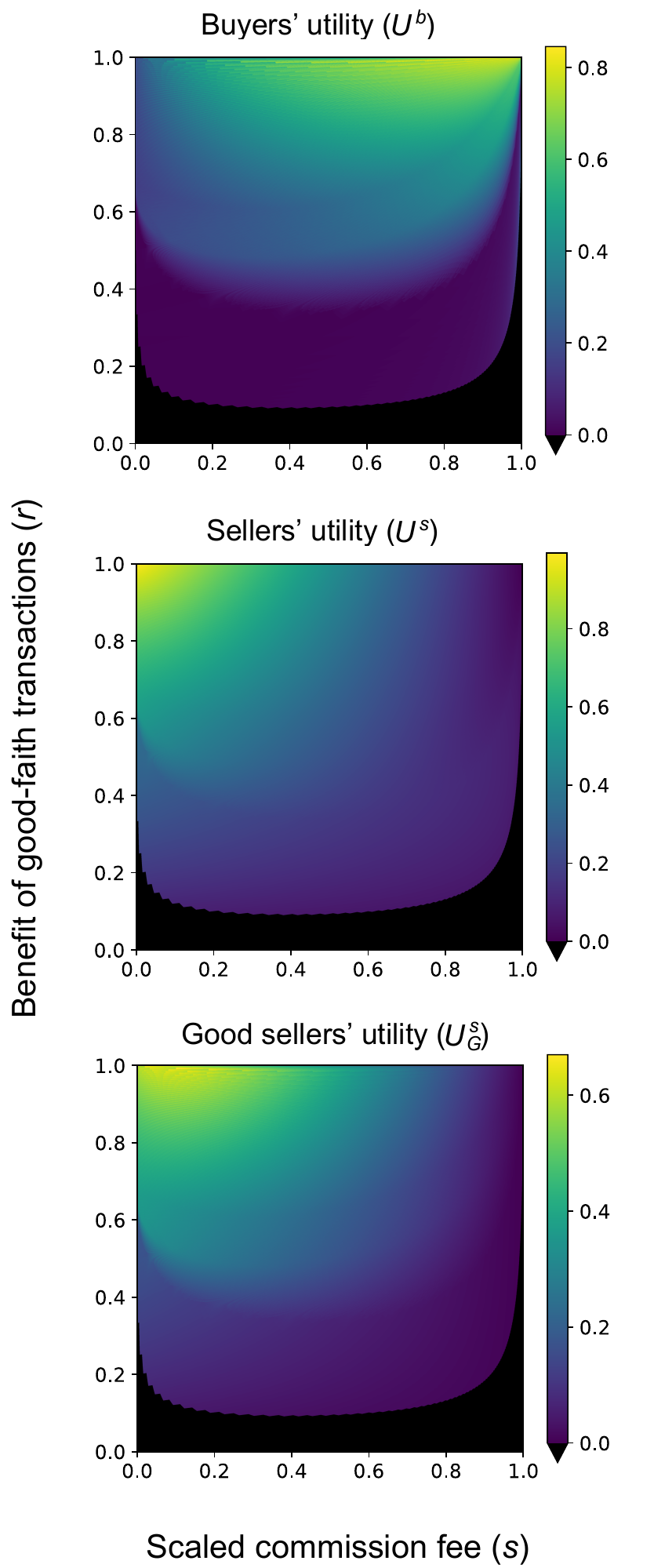}
\caption{\small Social utility produced by a platform for various values as a function of the scaled commission fee $s=c/r$. We plot the utility of buyers and sellers after the platform has optimized the true- and false-positive rates, $\alpha=\alpha^*$ and $\beta^*$. In this case, buyer's incentives (top) align more closely with the platform's incentives (compare top panel to main text Figure~5)
top), because higher commissions require more accurate reputational signals to keep trustworthy sellers in the market. As a result, buyer utility tends to increase with $c$ (top), whereas seller utility decreases (middle and bottom). In some regions no transactions occur at all (black areas) because the platform does not invest sufficiently in accuracy. Parameters: $\kappa=0.2$.}
\label{fig:ub_us_usg_rs}
\end{figure}

\clearpage

\begin{figure}[H]
\centering
\includegraphics[width=0.8\linewidth]{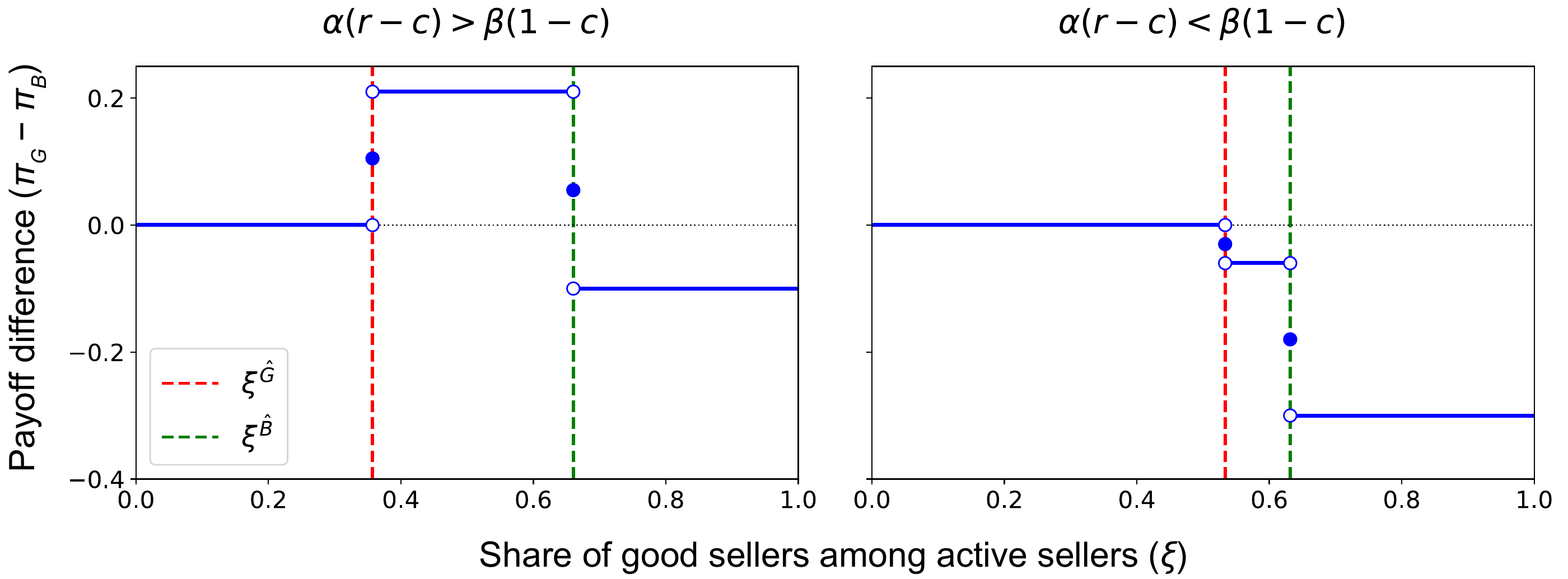}
\caption{\small Payoff difference between good and bad sellers. When $\alpha(r-c)>\beta(1-c)$, good sellers have advantage when $\xi^{\hat{G}}<\xi<\xi^{\hat{B}}$ and, thus, $\xi$ increases. On the other hand, it decreases in this region when $\alpha(r-c)<\beta(1-c)$. Beyond $\xi^{\hat{B}}$, bad sellers are always advantageous, leading to a decrease in $\xi$, and the dynamics is driven by neutral drift below $\xi^{\hat{G}}$ where no transaction occurs. Parameters in the left panel: $r=0.9, c=0.1, \alpha=0.6, \beta=0.3$. Parameters in the right panel: $r=0.7, c=0.1, \alpha=0.5, \beta=0.4$.}  \label{fig:payoff_diff}
\end{figure}

\end{document}